# THE PLANETARY MACHINE DESIGNED BY JOHANNES KEPLER

DANIELE L. R. MARINI[1]

ABSTRACT[2]

While Kepler was still working in Graz during 1598, some letters to his mentor Michael Mästlin demonstrate his interest in astronomical clocks and machines.
The first letter, dated January 6th 1598 contains a detailed description of a machine. In the second letter, written between June 1st and 11th 1598, Kepler starts with a brief review of clocks and machines of his time, then goes on to describe the requirements necessary for a useful mechanical instrument, based on the latest information of the day. In the *Epitome Astronomiae Copernicanae* (1618) he reiterates the importance and utility of astronomical and horological machines to divulgate the Copernican model of the Cosmo, to inform and assist scientists in their celestial calculations and hypotheses, even during periods of poor visibility in the night sky. I will present a translation of the Kepler's design and a hypothetical three-dimensional virtual reconstruction of his machine. This project reveals Kepler's ongoing research and understanding of the planets' cinematics, still bound to the homocentric spheres concept while the idea of *orbita* was maturing. At the same time Kepler's project reveals a reasoning on a clear description of retrograde motion of planets, fully developed later in his *Astronomia Nova*. His machine demonstrates the Copernican concept of the Sun and its planets as a unique system. He also wants to show how the planet moves from the viewpoint of an Earth based observer. He shows how to solve the basic mechanical problem of moving all the planets simultaneously with just one driving mechanism, which was impossible to accomplish with the Aristotelian theory of homocentric spheres.

## INTRODUCTION

The origin of astronomical machine is a long story that dates back to I century BCE, the Antikythera machine. Some of the ancient machines were mainly tools for observing the sky (armillary sphere, quadrant, astrolabe …), others were devised as a medium to explain the motion of the planets, the moon and the sun. Eudoxus with the mathematical abstraction of homocentric spheres, Apollonius and Hipparchus with the theory of epicycles and finally Ptolemais with the introduction of the equant, were the scientist that developed astronomy until I century CE. Thirteen centuries later astronomers and mathematicians like Georg Peurbach, revamped the study of these theories, until Nicolaus Copernicus, placing the Sun in the center of the Cosmos, set the ground for the search of a physical and dynamical description of the solar System.

This new image of the Cosmos was difficult to be accepted for religious and rational reasons. Johannes Kepler in his work *Epitome Astronomiae Copernicanae*[3], published in 1618, wrote that *Automata Coelestia* could be useful tools for teaching, for aiding the calculus and to help reasoning about the sky.

Kepler in 1596 exposed the idea of platonic solids in his opus *Mysterium Cosmographicum*, recognizing the signs of the Creation in the harmony of mathematical principles. To better describe his invention, he proposed to the Duke of Württemberg to build a model with the shape of a cup and the elegance and accuracy of a luxury object. The work on the Cup was long and complex, a Penelope's loom in the words of Kepler[4], due to the scarce interest of the goldsmith in charge and of the complexity of the work. Therefore, Kepler proposed to create a different machine to show the motion of the planets: a planetarium. In a letter to his mentor Michael Mästlin, Kepler describes the functions of the machine, that will display the motion of the planets around the Sun, with their proper periods, and positioned at a distance in scale as determined by his platonic solids model.

---

[1] Dipartimento di Informatica "Giovanni Degli Antoni" (prior to retirement). daniele.marini@unimi.it, danielelr.marini@gmail.com
[2] This paper was presented at the XLI Scientific Instruments Symposium, 17-23 September, Athens
[3] (Kepler, Epitome Astronomiae Copernicanae 1618) p. 26 (p. 7 of the transcribed manuscript)
[4] Letter 89, 1598, (Caspar, Johannes Kepler Gesammelte Werke 1945) «*Transeo ad Penelopes telam*»

Scientists and clockmakers of his time, except Wilhelm Schickard as we will see below, did not consider the project of the planetary machine by Kepler, who wrote only letters and did not publish anything else about this subject.

Walther von Dyck[5] published a transcription of January 1598 letter to Mästlin; he writes that his first objective is to show the common idea of the universe as if observed from outside but with bodies moving as the Copernican theory. His purpose is to explain to lay people the Copernican concept of the Solar System. The motion of the planets shall be displayed free, without obstacles that obscure the vision. Moreover, Kepler wants to show the motion of the planets around the Sun by keeping fixed the position of the Earth, to demonstrate that the Copernican idea does not conflict with the apparent motion.

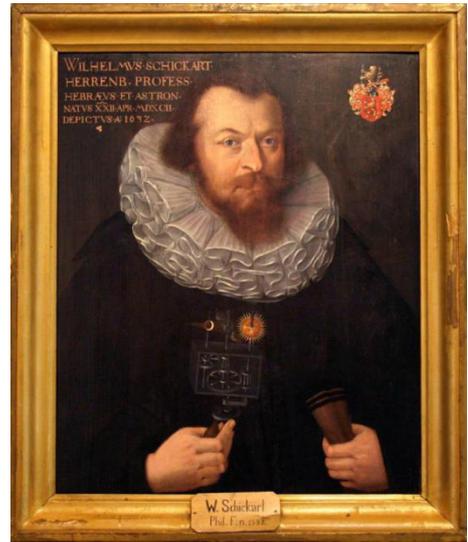

*Figure 1 - Portrait of Wilhelm Schickard*

The synthetic description of the model by van Dyck is close to my interpretation and virtual reconstruction, the major difference is that van Dyck does not discuss about the orthogonal assembly of the driving and driven gears, as we will see. He does not acknowledge the difficulty of assembling the system to allow the rotation of the whole globus around the fixed position of the Earth. In the background of Kepler's project, van Dyck glimpse correctly the theory of platonic solid. About the eccentric motion of the planets, Van Dyck does not explain Kepler's idea of using a flexible arm to support the planets that allows to slide on the surface of deformed spheres; he simply says that Kepler give some hints to solve this issue. Van Dyck compares the periods computed by Kepler's gearing scheme to those currently known and records the small differences. He also compares Kepler's periods to those computed one century later by Christiaan Huygens who used the method of continuous fractions, while the method adopted by Kepler is unknown and he probably got the values by trial and error. Van Dyck recalls also the great difficulties encountered by Kepler for the construction of the wheels and the limits of the ability of the goldsmith.

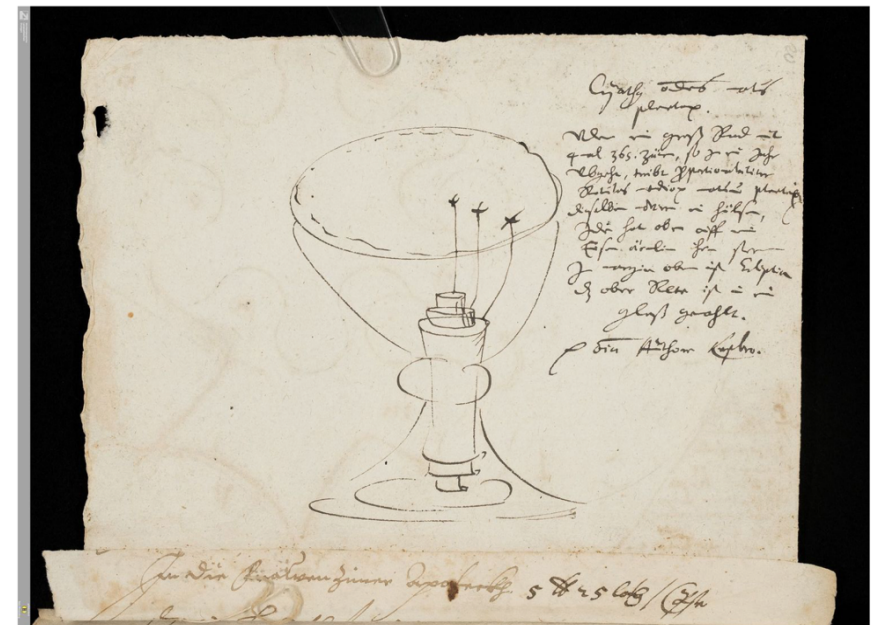

*Figure 2 - The sketch by Schickard*

Frank D. Prager [6] studies the work of Kepler as an inventor, considering his works about gear mechanisms, in particular for a hydraulic pump and for the planetarium. An interesting point noted by Prager, is the attention to technical drawings: Kepler appears to use isometric axonometry in the drawing of the Cup in *Mysterium Cosmographicum*[7]. It is anyway questionable Kepler's drawing ability, since his sketches of the planetarium are far from having the required technical clearness, compared to the drawing ability of artists of his time. Prager resumes the history of Kepler's project, from a *Kredenzbecher* to the Planetarium, and recalls the many difficulties encountered during this work. Prager recalls that Kepler was creating a mechanism since 1592, when he was just 21 years old, but no other drawing or sketches have been found earlier than 1598[8]. The key point in Kepler's planetarium, writes Prager, instead to show the Aristotelian homocentric sphere, is the will to create a model with a first mechanism to move each planet in a heliocentric configuration and at the same time reproducing his *Hauptinventum* – the platonic solids theory - and a second mechanism (coupled with the first) to show an earth-centered motion. Prager identifies in sketches 1 and 4 (see below) the outline of this idea, that I consider better explained with sketch 7. In his paper Prager recalls that Hans Kretzmeyer and Heinrich Schickard judged impossible to cut small wheels with as many teeth as 344. Wilhelm Schickard (1592-1635), nephew of Heinrich, mathematician and mechanic who also invented a calculating machine, considered to make a machine from Kepler's concept. During 1617 and 1618 Kepler and Shickard exchanged letters, trying to reach the objective to offer to lay people a three-dimensional model of the solar system. A sketch by Schickard is reproduced in Prager's paper, that clearly outlines the general shape of the model (see figure 2)[9]. The notes on the sheet have been interpreted by Prager as «*Cythii anders als planetarium (Another form of the Cup as Planetarium) On a great wheel with 4 times 365 teeth, so it goes for two years, moves proportionally rotating in radius(?) the planets. These planets as consequence of tubes, each over an iron arm. In the margin over is the Ecliptic. The upper Rete (?) is painted on glass. Pro Sin Authore Keplero*»[10]. In figure 1 we can see a portrait of Schickard holding a small planetarium, more properly a tellurium, that differs from Kepler's concept.

---

[5] (von Dyck 1934)
[6] (Prager 1971)
[7] (Prager 1971) p. 385. This is an interesting point also in my view: he use of *graphicum* in the title of his work reveals Kepler's attention to representation technique. At the time *graphice* was used «... *about the* **perspective** *when things are transposed and outlined on a plane from a long or short distance* » (Vitali 1559) p.311
[8] A drawing of a first project of the *Kredenzbecher* is in a letter to the Duke Friederich dated 1596. See below.

Henry King[11] wrote a brief description of Kepler's machine. Quoting his words: «*... he proposed representing the orbit of a planet not by a section through a sphere but by the path traced by the end of a mobile arm. The latter, radial and curved upwards, was fixed to a tube mounted on a central sun-stem. The central assembly, mounted inside the lower half of a sphere, therefore consisted of a set of coaxial tubes rotated by wheelwork at their lower end. His proposed teeth numbers are listed ..., from which it will be evident that the wheels were arranged in two parallel arrays or stack. Each tube-wheel was driven by a single wheel mounted on a vertical annual-arbor which also carried five other drive-wheels. Wheel 11 on top of the annual-arbor meshed with wheel 324 on the tube of Saturn, wheel 29 actuated wheel 344 of the tube of Jupiter, and so on down the stack. Complications arose when Kepler proposed reproducing a planet's change in heliocentric latitude and also the difference between its heliocentric distance at aphelion and perihelion. To achieve this first he considered incorporating a cam for varying the tilt of a planet arm. For the second he suggested making the arm's length variable or extendable so that it could follow the contour of a surface mounted eccentric to the model sun.*».

I did not find any reference in Kepler's letter to a cam to control the variation of the orbital plane, and for the eccentricity Kepler proposed a flexible arm that slides the planet on the internal surface of the sphere deformed into an oval shape. Also, H. King does not mention the first orthogonal disposition of the gears, as well as the concept of the primum mobile and the disposition of the solar system in the globus centered to the Earth, while all bodies rotate around the Sun.

Adam Mosley[12] explores the role of 'mathematical' instruments as computational aids and collectibles for aristocracy. This is the purpose of the two projects of Kepler, the first as a static sumptuous object for entertaining the Duke and his guests that could use it as a container of different beverages in the spheres, to be poured through separate pipes, while displaying at the same time Kepler's invention of the platonic solids to describe the solar system. The second project, started after the failure of the first, as a moving planetarium to be used either to better describe the Copernican model and to compute and emulate planetary motion. Mosley does not enter into details of the project, and emphasizes the opportunity for Kepler to create a rich and complex machine to disseminate the theory of Copernican world system. Mosley recalls the overview of existing astronomical clocks and machines described to Mästlin by Kepler, from which it is evident how the idea of Kepler was different from objects capable of displaying only the average motion of the planets on separate dials. His thought was to embody the geometry of the cosmos to obtain a cosmologically and mathematically accurate representation.

Rhonda Martens[13] discusses the role of models and representation in Kepler's scientific methodology, and underlines the importance that model shall be strictly linked to a mathematical description of the reality. Martens concludes that the preferred medium to Kepler were machines and three-dimensional models rather than diagrams. This explains the large use of drawings and the concept of the planetarium. I can describe this result saying that Kepler was visually thinking, he was imaging himself as observing the solar system from an external viewpoint, resulting sometimes in the concept of a solid three-dimensional model and sometimes ad a plane diagram of planet's motion, thus capturing the mathematical nature of the astronomical phenomena.

---

[9] (Schickard 1610)
[10] Author translation
[11] (King e Millburn 1978) p.92
[12] (Mosley 2006)
[13] (Martens 2009)

Noam Andrews [14] analyses the first Kepler's idea of the Cup, the *Kredenzbecher*, that, before moving to the construction of a *Globe* with planetarium, was discussed with the duke Friederich and with Mästlin. A drawing of this first concept is included in the letter 26, February 17th (see figure 3, from Andrews, cit.). The first concept was not to build a model of the table in *Mysterium Cosmographicum*, rather to create a luxury object for the *Kunstakammer* of the Duke: a cup made with nested spheres of gold or silver, sized and distanced following the platonic solid theory. A complex set of tubes will be constructed to allow to spill wine or liquors contained in the different spheres. The research by Andrews is focused on the difficulties of the construction of the *Kredenzbecher*, and he does not study in details the planetarium project. Anyway, Andrews identifies the inability of Kepler to lead a complex construction program as one of the reasons of failure.

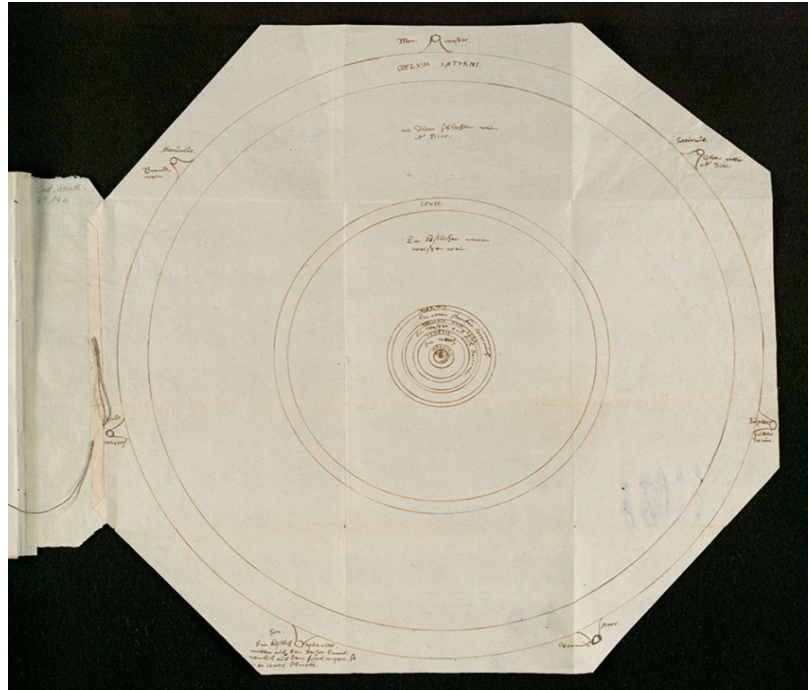

*Figure 3 - Drawing of the Kredenzbecher by Kepler. Letter 28*

In this paper I will present a translation of the letter of January 6th 1598 to Mästlin about the project, commenting and interpreting many obscures sentences, and I will summarize a letter to Mästlin, written on June 1-11th the same year, where Kepler overviews known astronomical clocks and automata, and provides a better description of the aims of his project. In order to appreciate the burden and complexity of the work undertaken by Kepler, I will also recall key moments, as documented into letters among Kepler, Mästlin and the Duke. I will also present and discuss a three-dimensional virtual reconstruction of the machine, based on my interpretation.

### *KEPLER'S PROJECT*

The project of the astronomical machine is described with details by Kepler in the letter to Mästlin[15], January 6th 1598, whose relevant parts are translated below.

I could not access the original manuscript, anyway I considered two transcriptions: 1) Kepler Gesammelte Werke volume 13, edited by Max Caspar, containing letters from 1590 to 1599, among them letter number 85 that includes Kepler's sketches of the project; 2) the transcription published by Walther von Dyck[16] in 1934 where the drawings have a different look.

Kepler's letter is written in a mixture of Latin and old German. I did not translate the term *"orbis"* (declined as necessary), the possible translation *orbit* is not suited, since this word has been introduced years later in

---

[14] (Andrews 2021)
[15] (Caspar, Johannes Kepler Gesammelte Werke 1945) p 162-176
[16] (von Dyck 1934)

*Atronomia Nova,* (1609)[17]. I did not translate other some Latin terms that were customary in the scientific literature of the time. The page and line numbering are from Caspar's edition.

The fragment from page 163[18] line 26 to line 51 introduces the subject:

Ut autem mihi adsim in hoc negocio, scribam tibi consilium totius fabricae. Instrumentum meditabar primi et secundorum mobilium simul et una demonstrationem mei inventi. Ut hoc obtinerem, illa expressu difficilia fuerunt addenda. Primum, das man alle orbes vnd 5 corpora solle abheben thönden. Mitt den orbibus hatt es khein not, dan sie werden mitten von einander getheillt. Mitt den corporibus aber hab Ich wöllen kunst brauchen, vnd also anordnen, daß Inen im abheben khein beyn (σκελος sive latus) zerbrochen oder zertheilt werde. Auch hab ich vermeindt, es werde khein Kunst sein wan man nicht alle orbes vnd corpora auß einander lösen vnd gleich als ein Uhr zerlegen thönde. Wan es aber je zu schwehr wollt werden, So thue man eins, man mach alle corpora vnd orbes (vom kleinesten anzufahen) in einander vnd löt oder niette sie zusamen, daß sie nicht zerlegt werden thönden. Danach feile man das gantze opus mitten entzwey, daß es nun zwey stuck gebe vnd setze dan circulos planos repraesentantes viam planetarum füeglich darein, das man es also zustürtze thönde (ad speculationem mei inventi) oder abstürtze, wan man es pro Theoria brauchen will. Zum andern, hab Ich zu behauptung des namens (Theoria primi et secundorum mobilium) verordnen müessen, das das gantze opus internum, so da durchsichtig ist, auff vier pfosten in medio ♉ ♌ ♏ ♒ angeleinet berumb getriben werde. Damit wan man gradum solis vel terrae in circulo terrae (beneficio circuli per medium circulum saturnium sculpti, et aequaliter divisi) signirt, man alsdan das Werckh ruckhe biß derselbe signirte gradus in das mittel globi stellati khomme.

*Moreover, to continue about this issue, I am writing to you for an advice about the whole construction. I was considering to build an instrument to demonstrate my invention and at the same time the primum and second mobiles. To get this already difficult result, I had to solve further difficulties. Firstly, all the orbes and the five bodies [planets] must appear raised and standing out. The orbes will be split in half not to obstruct each other. Regarding the bodies I had to devise a solution and arrange them so that their supports (σκελος leg[19]) are not broken or split. In order to achieve an artistic work, I also made sure that they do not intertwine and that they can be dismantled like clockwork. If this proves too difficult, all the orbes and bodies (from the smallest to the largest) will be fastened together with rivets. The entire opus is therefore divided in the middle into two parts, so that the flat circles representing the paths [viam planetariam] are inserted (in order to be able to examine my work) or removed in order to use them to show the Theory. Furthermore, to justify the name (Theoria primi et secundorum mobilium), I had to arrange the inside of the entire work transparently so that one can look through and see them dragged through the middle of the zodiac signs ♉ ♌ ♏ ♒ supported by four supports. In this way one can mark the position in degrees of the Sun or the Earth on the circle of the Earth (constructed in the same way and engraved regularly as the circle of Saturn) and make the machine march until it reaches the same sign in degrees in the middle of the starry globe.*

From this short description we can figure out a general view of the machine: a globe divided into two parts, inside which the planets and orbs can move and show their motion. The machine, if possible, could be disassembled. It can be used to examine his theory by inserting circles that display the planetary paths, or to show his theory, by removing the circles. The globe will be engraved with the position of the stars; along the planets and zodiac circles are marked and engraved divisions into degrees.

Kepler describes the procedure for engraving, but during the construction the goldsmith made a huge «*immanem*» error: the Saturn circle was divided into 396 degrees instead of 360! He thinks he can rectify the error, but he understands that the goldsmith is not a professional engraver. He considers the possibility to give back the silver to the Duke, to apologize and ask to buy new silver, but the work already cost 1.000 florin. If manufactured in Augsburg, Nuremberg or Antorff (Anvers) the cost would be ten times less. Kepler considers to move the work to another town with more professional artisans, and hope to get the authorization of the Duke. In the meantime, he examines three options: to complete the opus with the defect, or to give up and return the silver, and finally to stop the current work and start again from scratch, but risking new problems. Kepler would choose the third option, but the silver given to the goldsmith has lost the fineness of its alloy, for multiple melting, and this could cause litigations with the goldsmith. He will need more silver and support other costs. He is also very concerned about associating his name to a failure.

From page 171 Kepler describes the mechanism of his project[20].

---

[17] (Goldstein e Hon 2005)
[18] Manuscript pages: 219v, 220
[19] Translation by Kepler
[20] All figures are by Kepler

*I reflect day and night, when it comes into my hands, how I can do better. I really want to write about my result of how motion begins, because I have been thinking about it for six year. Like this: each planet must move as in the real world (excluding the Moon, it is too small). There will be no special pointer moving and leaving the Earth, because that is impossible, it could interfere with the others. But a pointer must be fixed to the Earth, which each person can apply at will to the body of a planet, to show where it is in relation to the Zodiac. Or it can be placed sometimes at this, sometimes at another planet, so that one can see how each one moves retrograde, etc.*

 *The internal mechanism, as I have already observed, moves with non-circular motion, but like an epicycle that preserves the eccentricity of the planet. Thus, to make each part rotate, it is fixed on a rigid support in the form of an arc AB whose extreme A goes into the "mobile" C and the other extreme B goes under the foot D [sketch1]. In this way it can be moved rapidly for a year, or if you wish by the mechanism of a clock. Care must be taken, however, that the hole B does not pass through the center of the mechanism or the Antarctic pole of the Zodiac.*

*Now I clarify better the hook AB. In A there are six discs stacked so that BA rotates in B and pulls A. The upper discs are small and the lower large, and they have equally spaced teeth [with the same pitch]. Marking many teeth on a small wheel is difficult (for the others [larger] it is easy). But two things are to our advantage: 1. The teeth will not be overloaded, they do not lift but merely rotate. In fact, the load bears on E (and is partly suspended on the inner section C of the opus with the Zodiac), as we will shall see later [Sketch 2 and 3]. 2. If we choose the size of the fixed star sphere arbitrarily, there are not constraint on the silver. Let us proceed. CE consists of 6 concentric tubes, the outermost short, the innermost ones gradually longer and more protruding (NB: the wheels in A are not straight, but lie flat, so that CE is perpendicular). An A-wheel is mounted at each end so that the longer A-wheel is longer and the inner A-wheel is shorter. I cannot draw: you must try to understand [«non possum pingere, tu intellige»]. The greatest effort to drag the motion is here, since the shells rotate enclosed within each other. the rotation does not require a great force because the tubes are enclosed within each other in fact they do not unload their weight on each other. The load weighs on E as much as one wants, nevertheless it weighs but separately, as we shall see. And if the weights weigh mutually on each other, nevertheless they will be light once the bodies and orbs are removed.*

*Now it is time for the number of teeth. (Note: I have many possible sets of numbers, here I only put one [Tab.1], not because it is more suitable than another for construction, but because the calculation is less difficult). The wheels in E have different teeth, separated at the centre of the equant into four upper ones [Earth, Mars, Jupiter, Saturn and two lower ones Venus and Mercury]. Regarding ☿ [Mercury] the error relative to the mean motion during a 10,000-year period is 290*

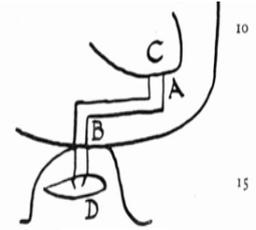

Sketch 1

Ich tracht im Tag vnd nacht nach, wie Ich, wans mir vnder die Hände khompt, auffs allerbeste mache. Wolt gern meinen anschlag de motu schreiben, bin schon sechs Jahr mit vmbgangen. Doch soviel:

Es soll ein jeder planet lauffen, wie er in mundo lauft (excepta luna, die ist zu klein). Khein besonderer Zeiger soll mit lauffen, qui ex terra prodeat, dan das ist Vnmüglich, sie hindern einander. Aber ein Zeiger sol dabey sein, affixus in terra, den ein Jeder pro arbitrio corpori planetae appliciren mög, vnd sehen, wa er im Zodiaco stehe. Oder zuweilen disem zuweilen jenem planeten affigirn, damit man sehen khönde Wie es in retrogradationibus etc. uniuscujusque zugehe.

Das innere Werckh, ut antea nosti, movebitur non conversum, instar epicycli alicujus quo planetae eccentricitas salvatur: ita ut semper easdem partes iisdem fixis obvertat, das geschicht durch einen starcken trib oder Hacken AB, da der eine Hack A yber sich gehet in die mobilia C, der ander B vnder sich durch den Fueß D. Da kan man in nun eintweder treiben, das in eim augenblick ein Jahr vmbgebe, oder kan in, wie folgen soll, an ein Taguhr befften vnd lauffen lassen. Es mueß aber das loch B auch nit mitten im Werckh oder in polo antarctico Zodiaci durchgeben auß Vrsachen die folgen werden.

Jetz will Ich den Hacken BA besser erklären. A seind sechs blat auffeinander befft vnd gebt deren theils vmb nisi qua BA in B vertens, ipsum etiam A vehit. Die obere sein klein die vndere groß vnd breitt, die haben Zän aequaliter divisos. Da ist maxima difficultas mechanica (caetera facilia) vil zän in ein klein rad zubringen. Sed duo commoda. 1. Die zän dersten nit starckh treiben, non sursum, tantum circum. Nam pondus incumbit in E (et partim pendet in sectione interioris operis C cum Zodiaco), ut audiemus.

2. amplitudo fixarum si liberemur ab argentea materia ist arbitraria. Sed pergo: CE seind Hülsen, sechsfach yber einander (NB. die räder A stehen nicht also aufrecht, sondern ligen, ut iis CE fiat perpendicularis) die äußere kurt, die innere länger, prominentes. In extremitate cujusque ein rad auff die räder A gerichtet, also daß superioris et exterioris arundinis circulus groß ist, interioris klein. Non possum pingere, tu intellige. Die größeste gewalt der Trib stehet darin, das die Hülsen circum vbereinander schließen. Nam non sibi mutuo

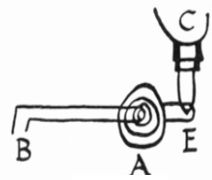
Sketch 1

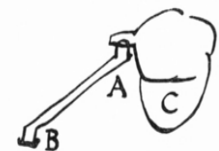
Sketch 3

incumbunt. Quamvis enim omne pondus in E incumbat, tamen magis pendet, et singula quidem separatim, ut audiemus. Et quamvis incumberent sibi mutuo omnia, tamen levissima sunt: demptis corporibus et orbibus qui pendent. Nun die außtheilung betreffend helt sichs die also (habeo multas formas numerorum, hic unam tantum pono, non commodiorem artifici sed minus vitiosam in calculo)

| | | Tab. 1 | | | | Anni 10000 |
|---|---|---|---|---|---|---|
| | oberst 1 | | 11 | 1 | 324 | 88 |
| Das | 2 | rad | 29 | 2 rad | 344 | 600 |
| | 3 | in | 42 | Das 3 in | 79 | 139 de |
| | 4 | A | 60 | 4 E | 60 | 0 |
| | 5 | bat | 395 | 5 bat | 243 | 23 de |
| | vndst 6 | | 191 | vndst 6 | 46 | 290 de |

Aber die Räder in E haben vngleiche zän, divisos ex centro aequantis in 4 superioribus. Defectus in ☿ omnis 10000 est 290 gradus in motu medio, annis 1000 est 29, annis 100 est 3° fere. Sic in caeteris.

*degrees, 29 degrees during 1,000 years and about 3°
during 100 years. This is also the case for the other
planets.*

*I have already written that opus C, suspended in its centre
from the outer part of the moving sphere of fixed stars, with,
try to understand, balanced supports, can be difficult to
construct. C is led by E, which rotates in the opposite direction
around B, preserving the fixed parts. [Sketch 3]. Similarly
BE is bent around A so that while AE and EB rotate, A
rotates B concentrically and E eccentrically. I also wrote that
the hole in B should not be in the pole of the Zodiac, but as
far from it as the length of AE. This distance determines the
size of the work and how to make the wheels, which must be
as small as possible. In this way, the A-wheels will be fixed to
their axle and the E-wheels will be mobile and rotate freely
(with the E-wheels reversing their motion in relation to the A-
wheels). Opus C, as explained, is conducted so that the Earth
remains at the center of the orbis of the fixed stars. It is clear
how beautifully these motions are conducted. It is evident
that if the motion of the Earth in C takes place along HIK, an
opposite motion KIH is needed to keep it in the center [Sketch
4]. For if the Earth moves correctly, the other planets will
also move in agreement in the sky, and together with the
Earth.*

*Now let's come to C. It will be made with orbibus divided in
half and without holes. They will be fixed to the bodies so
that no orbis will be movable.*

*Saturn's outermost tube will be the shortest of all and goes
free in the outer surface of Saturn's shell to hole PQ [Sketch
5], and here it closes but has a Y-hook that rotates between
the two surfaces of Saturn's shell, and on the top of Y is the
planet. I have already explained above by which mechanism
the planets move slow or fast.*

*I will now describe how they ascend or descend. On the
external surface of Saturn's hull [orbis] there is a convexity
in the aphelion, and a depression in the perihelion so that Y,
like a thin watch hand runs lightly over the surface on which
it presses and is repelled like a spring. In this way the motion
of the four superior planets could be shown perfectly. Could
I solve the variation of latitude easily in my own orbis? Or
how else could I do it? If I had an inclined surface standing
free under the tube of Saturn, it would move Y up and down.
But also, the tube of ♃ [Jupiter] with all its two parts crosses
the orbis of ♄ [Saturn] and the center of the cube plane [linea
cubi] enclosed in the orbis of ♃. A rod crosses the orbis of ☿
[Mercury] up to the sun. I have not yet thought about
variations of the motion of ♀ [Venus] and ☿. ♀ Variations
of ♀ are not necessary because the opus is small; for ☿ there
is no difficulty because its motion is hidden between the
surfaces of the orbis. I could solve it with a spring-loaded
compass that extends and contracts.*

*There is no need to say anything about the Moon because
its motion is uniform. I could use the same trick as ☿, but I
haven't thought about it yet. I want to determine and fix the
eccentricity, if I can, at 1/10 of the diameter everywhere
(since in my invention, as I have already described, the
calculation changes what is equivalent). Similarly, I can*

Dixi opus C in medietate sui ab exteriore ambeunte sphaera fixarum pendere, id intellige mit ebenmäßigen umbgebenden Hacken, praestat fortasse arduis. Per eos fit ut opus C vectum ab E (in altera facie) circa B semper ad easdem fixas iisdem partibus respiciat. Itaque et BE in A flectetur ut alternis AE cum EB coincidat vel in rectam lineam dirigatur, et A describat ipsi B concentricum, E eccentricum. Ideo dixi B foramen non debere in polo Zodiaci esse, sed tantum ab eo distans quanta est amplitudo AE. Illam autem amplitudinem definiet artifex, dentibus in rotas fabricandis, quam possunt fieri minimas. Hoc modo fit (cum quae in E rotae, easdem iisdem partibus fixas in compositione respiciant, quae vero in A, fixas percurrant) ut rotae A immobiles, rotas E mobiles moveant in partem oppositam. Vehitur autem opus C, ut nosti ideo, ut terra per omnem suum ambitum maneat in centro orbis fixarum. Unde patet, quod hi motus pulchre invicem aptentur. Videlicet quia terrae motus est in opere C per HIK, oportet illi obviam iri motum alium, qui illam retineat in centro: qui est quasi KIH. Quia igitur terra recte movetur, ergo movebuntur et reliquae planetae in convenientem plagam. Omnes enim cum terra in eandem.

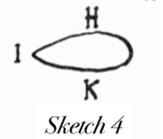

*Sketch 4*

Sed ad ipsum opus C veniamus. Id facerem dimidiatum, ex orbibus binarum superficierum solidarum, non pertusarum. Illi per corpora sibi mutuo affigerentur, ut nullus orbis esset mobilis.

Die eüssere Hülsen Saturni waere utrinque am kürtzesten, die gieng in exteriore superficie Saturni im loch PQ frey, das ist, sie verschlöße sich vnden in Saturnum vnd da börete sie auff, allein das sie hett einen zeiger Y, der sich zwischen den zweyen superficiebus Saturni vmbkherete, vnd oben in Y hett er den Planeten. Dixi supra quo machinamento planeta in eodem ambitu tardus vel velox fiat.

Iam quomodo ascendat vel descendat dicam. Es muest im vmbkreiß superficiej ♄ ein bausch sein der in in ἀφηλίῳ auß, in περιηλίῳ ein trieb, der zeiger aber Y, der vn das trump, müest feder waich sein, das sich leicht druckhen ließ, vnd wider zu sich selbs schnapete. Hoc modo rectissime omnis motus diversitas in 4 superioribus exprimeretur. Khöndt im auch leicht in latitudine helffen, si ea simplex sit in ipso orbe, sed ad quid? Geschehe also wan die hülse Saturni, die da frey gehet, vnden da sie aufsligt, einen vngleichen Weg hette. Da gieng dan Y hoch vnd nider. Nun also fortan die hülse ♃ mit allem jrem inwendigem güeng per orbem ♄, per centrum plani cubicj hinauff in orbem ♃, vnd also fortan, oben in medio orbis ♃ güeng der längeste stilus von vnden biß oben auff vnd trüeg die sonne. De ♀ et ☿ varietatibus nondum cogitavi, in ♀ parva est necessitas, quia opus parvum, in ☿ verò nulla difficultas, cum habeam motum et quietem unà, deinde et latebras inter superficies orbium. Khöndt ime mit eim federcirckel helffen der sich auß vnd einzöge. De Luna plus necesse non est ut exprimatur, quam ejus aequalis motus. Id fieri potest ijsdem commoditatibus, quae sunt in ☿, quamvis nec de illâ adhuc cogitarim. Ich will mich drüber setzen vnd raitten (· quandoquidem ipsum meum inventum, quod par est, in opere exprimi, calculum anteà mutat·) ob es vil außtrage, wan ich ubique $\frac{1}{10}$ diametrj pro eccentricitate näme. Item, wan ich ♀j sein apogaeum liesse, wie es ♄, ♃, ♂, ♁ hatt. Quaeris, wie ich woll superficiem exteriorem vnd interiorem an einander henden, si intra omnia libera esse oportet? Difficilis quaestio. Si dicam per circu¹lum, qui OS latitudinem habeat: ubi planeta ibit? Si brachiolis dicam, oporteret planetam habere latitudinem, et brachiola fieri non nisi in una medietate. Si supra planetae viam superficies extendam et ibi connectam: jam via clausa erit volvendae regulae. Itaque existimo non multum incommodaturum, si machina paulò admodum sit altior dimidio globo et planetae sub tribus vel 4 brachiolis ceu sub jugum missi, transeant. Latitudo tamen planetarum addita, si tanta esset materiae firmitas, rectius juvaret, si nempe lingulae ABC, quae non *leave the apogee of ♃, ♄, ♂, ☉ as it is. You ask how I want to tie the inner and outer surfaces together if they need to be free? Difficult question. If I answered: enough so that OS space is maintained along the rim, then where would the planet go? If I answered: on its Y arm then the planet would change its latitude, and the arm would have to maintain its median position. If I extended and connected the surfaces above the planets path then this would be blocked. However, I think it would not be too difficult if the machine were taller than the half globe and the planets, as if under a yoke, passed under three or four small rods [Sketch 6]. However, having solved the variation in latitude of the planets in this way, we have to support heavy material, and it would help if the ABC rods, which cannot be too large, also supported this weight. Without this, everything must be constructed to be light. How will the tubes be guided by the planets and the outer surfaces of their shells? This cannot be solved with bodies touching the inner or outer surface, as we have done with eccentricity. The difficulty remains that the orbis between the two surfaces is not open all around. It would be as if one wanted to force the law of motion (which is not discretionary by some trick or αυτοματος) that both halves of the interior of the opus are the same, so that the upper part would be transparent the lower part solid. Otherwise one could draw this diversity in such a way that it could be divided into two similar parts with at the level of the Zodiac, from which the retrogradation could be observed. As for the 'primum motum', it need not be closed and hidden within the opus. But if the sphere of fixed stars is built open in the middle or constructed as a grid, the outer planets will move along the horizon. They will move around a pole, and the lower part around a foot, remaining on a small moving ring that will be connected to the annual motion. I have pondered so much that, to remove my doubts, a master mechanic from Augsburg could advise me.*

*The explanation I gave of basic motion is difficult. I want to clarify it further. Let there be a rod CD, at point B there is a rod BA [21] [Sketch 7] that rotates around the fixed point A and keeps CD always parallel to itself. Point B on rod CD describes a circle G with center at A, while all the others have different centers, such as E describing a circle with center F. Now it is understood that behind what is drawn is the sphere of fixed stars whose pole is (as is easily understood) the center. A will therefore be an extra [eccentric] center or pole. AB is the semi-diameter of the orbis of the Earth and is equal to FE. E is the pole of mobile orbis (or the Sun in the center of the mobile). Thus the circle around F led by E is equal to the orbis of the Earth and it is as if F were led along a circle around E. Now imagine 6 motor circles pivoted in B, and passing through a hole in the rod CD. In E there are 6 free wheels adapted to the above. Now let B rotate towards G, then E will rotate towards A and H will rotate on the axis towards G (NB: it would be better to reverse the whole motion from right to left.) Let there be a circle around E led by the adjacent circle in the*

possunt esse profundae, possent sustinere pondus in alteram partem pendens. Es mueß ohne das alles inwendige leicht sein. Wie wan die Hülse cum planeta etiam superficiem exteriorem mitführete? Khan nit sein propter corpora vel extra vel intra attingentia, item verderbte die eccentricitatem. Bleibt derowegen bey ermelter incommodität, das orbis zwischen beeden superficiebus nicht rings umbher offen stehe. Es wäre dan das man sich der regel (·quae aliàs arbitraria non ex artificio vel αυτόματος·) verwegen wolte, da hefftete man bede halbe Theil operis internj zusamen also, das das obere Theil durchsichtig, das undere solidum wäre. Oder gestaltete man auch hanc informitatem das es also in zwey drittheil an einander hüenge in eim drittheil aber gleichsam einen schnitt hette a Zodiaco factum, darinnen man durch anhefftung der regel einem die retrogradationes zeigen khöndte. Iam quòd ad motum primum, ille non esset necessarius clauso et occulto interno opere. Sed si sphaera fixarum unâ medietate effingeretur pertusa et pervia, sive cancellata: facilimè moveretur das eüssere wesen, daran der Horizon hangt. Moveretur circa unum polum, et inferius curtator circellus incumberet rotae moventj, die khöndt man underm fueß mit dem Jahrtrib leicht vereinigen. Sovil hab jchs bey mir selbs bedacht, ist mir khein zweiffel, ein werckmeister zu Augspurg würde mir einen bessern rhatt in eim oder anderm geben khönden.

Ich hab den grundmotum gar schwer angegeben. Will in besser erklären. Esto linea sive regula aliqua CD, in cujus puncto B vertitur cardo radij BA. Volvatur ille radius circa A immo¹bile, et vehat secum lineam CD, sic

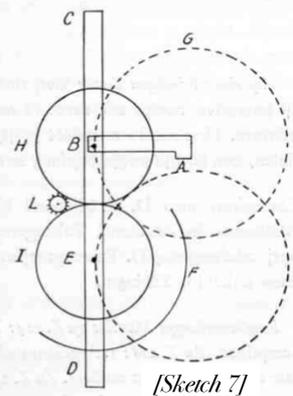
[Sketch 7]

ut illa semper sibi ipsi maneat parallela. Id si fiat, unicum punctum regulae ad B scribet circulum ad A concentricum, reliqua omnia eccentricos, exempli gratia E scribet circulum circa F centrum. Iam accommodatio: quicquid hic pictum est, intelligiatur in fundo globj fixarum. F esto polus fixarum, sive (·quod per intellectionem idem est·) centrum. A igitur erit extra centrum (·seu polum·). AB est semidiametro orbis terrestris aequalis, quae est FE. E esto polus mobilium (·seu Sol in centro mobilium·). Ergo circulus circa F per E ductus est aequalis orbi terreno, et perinde est, ac si F circa E circumduceretur. Fingantur jam circulj 6 motores ad cardinem B immobiliter affixi mobiles tamen (·ut cardo·) in foramine lineae sive regulae. In E sint aliae 6 rotae ab invicem solutae et aptatae ad illas priores. Fiat ascensus B in G. Ibi igitur E in A veniet, et H incidet in Regulam versus G. (*Randbemerkung:* Potius debuissent omnia convertj a dextrâ in sinistram.) Esset autem circulus circa E per se immobilis, jam verò urgetur a contiguo in contrarium, ut I in regulam incidat versus D. Qualis autem erit motus rotae inferius, talis etiam planetae superius qui adumbretur per KF, et sit terrae. In praesenti situ F terra est, E loco Solis. Fiat igitur per interjectam L rotulam aut aliter, ut H ascendente, I etiam ascendat. Ergo ubi E Sol venerit in A, veniet H versus C, I versus B, et sic F versus D, in lineam. Erit autem linea applicata ad centrum F. Terra igitur rursum erit in F hoc est manebit eodem in loco. Sic etiam de caeteris planetis. Illi enim propter inaequalitatem rotarum B et E, inaequaliter movebuntur, sive non simul redibunt. Sic 6 planetas 12 rotis et 6 actoribus (·sive ♎ ℔·) movebo. Potest autem BA radius extremitati C vel D, vel utrique adhiberj, ut tantò amplius sit spacium rotis multorum dentium. Das ist mein anschlag. Zum fordersten aber bitt jch, Cura ut te fatigarj meâ causâ dolere desinam: hoc est, ut te extrices et mihi domum remittas. Es solt mir wiß gott leid für euch. Deus tibi totique familiae, Collegio et Universitatj multa prospera largiatur. Valete et salvete omnes a me. 6. Ianuarij annj 1598 St. n. qui mihi dies annum 27 incipit. Gratij.

H. T. Gratissimus discipulus
M. Johan Kepler

*opposite direction towards I and D. Whatever the motion of the lower wheel, such will be that of the upper planet which is represented in KF[22], and that is the*

---

[21] The letters A and B in Sketch 2 and 3 are exchanged in Sketch 7

[22] Letter K does not exist, read as HF

*earth. In this situation F is the Earth, E the position of the Sun. An intermediate wheel L reverses the motion so that the Sun in E moves towards A, H moves towards C, point I towards B and F towards D. The Earth is and remains in F. This is also the case for the other planets. They, because of the different pairs of wheels in B and in E, will move with different velocities. Thus I will move 6 planets, 12 wheels and 6 actors. The length of BA can also be adjusted at the extremes C or D, so that there is enough space for wheels of many teeth. [The letter concludes with greetings]*

## KEPLER'S OVERVIEW OF ASTRONOMICAL MACHINES AND THE REQUIREMENTS OF THE MACHINE

Besides the description of the design, Kepler wrote also the requirements of his invention, compared with some of the most important astronomical machines known to him. In the letter to Mästlin, dated 1$^{st}$/ 11$^{th}$ June 1598 [23], Kepler answers to critics and request by Mästlin, and provides a description of the functions of the mechanical planetary that he is designing and constructing and an overview of astronomical clocks and machines.

The overview begins with a reference to Archimedes *sphaera*, as described by Cicerone in *De Republica*, that he considers the only effective representation of the cosmos. Kepler then quotes a machine by Posidonio, that "*with a single motor shows the motion of the Sun, of the Moon and the moving stars in the sky*". Kepler recalls that Ramus[24] lists two clocks in Paris, one in Sicily and one in Germany, all made with great art. Emperor Carl has a similar one that has been built probably by Francisco Turriano Cremonese[25]; this machine, as writes Cythraeus[26], should represent the motion of the planets with small wheels. The Landgrave of Kassel, Wilhelm IV, should have another similar machine, that Ramus calls *astrario*. A clockmaker in Dresden has seen another similar object, but it did not preserve the average motion of the planets, that is usually displayed separately. An astronomical machine is exposed in Münster, it does not have the average motion of the planets, that is shown on separate dials, anyway it is full of ornaments that "fill the eyes". The mathematician Jacobus Cuno[27], form Frankfurt am Oder, wrote a booklet to describe how a machine can be built that shows the motion of the planets with stations and retrogradation[28]. Kepler writes that in those time it was not possible to build a mechanism capable to display all the planets' motion driven by a unique wheel. He assumes that the many horological clocks, like that of the king Christian of Denmark, that has been donated to Moscow, could display only the average motion, like the tower astrolabes built in Nuremberg and Augsburg, that show the average motion of the Sun, of the Moon and the planets. Kepler writes that he has seen something similar, valued four thousand gulden. The young king of Poland, to celebrate the new year with pomp, asked to a mechanical master of Augsburg to build a machine that plays a music concert with 24 trumpets and a Bacchus and Silenus parade at each hour. Kepler admits that these are beautiful works of art, but he observes that they have only a celebratory function. Kepler mentions an astronomical clock that shows also the atmospheric weather, with snow and rain, a machine probably made by Johannes Stöffler (1452-1531). Paolo Giovio (1483–1552) describes a silver machine donated by Emperor Ferdinand to the Turk emperor Suleiman[29]. Cuno has spoken of the planetary motion represented with the variation of the latitude and longitude, with equinox precession and with all the information collected in the Prutenic Tables. All these motions should be represented with such accuracy that after six thousand years there would be less than a degree of error.

Kepler has perfectly clear that all the mathematicians that have worked on the mentioned machines could compute the position of 7 planets on a circle of 360 degrees, with the hours of the day on circles with 24 positions. He concludes his overview by observing[30]: "*it is not possible to move with a horological mechanism every planet, in the same way they flow in the sky with their proper distance*". "*Considering the Archimedes' sphere, he desired to show not the motion but the real proportions of the sky. … If in our time an artist would apply his art using only indicators on clock dials, he could make a sphere with a mechanism as well. Therefore, he could make with no doubts a mechanism where one could see how the orbes rotate around each other, as it happens in nature*". Kepler is well aware of the existence of many and various instruments created with diligence by machine makers with smart geometrical demonstrations: *«Hinc infinita varietas instrumentorum est orta, certantibus Mechanicis manuum sollertia, cum Geometrarum demonstrationibus ingeniosissimis».*[31]

---

[23] Letter 99, (Caspar, Johannes Kepler Gesammelte Werke 1945) p.218-232

[24] Giovanni Battista Ramusio (1485 – 1557). He was deeply interested in the geography and recent discoveries. As a member of the Venice administration, he was in contact with many scholars of his epoch.

[25] Geminello Torriani (c. 1500 – 1585), a builder of automata and an engineer.

[26] David Cythraeus, (1531 – 1600) a Lutheran theologian.

[27] Jacobus Cuno (1526 – 1583/84), mathematician and astronomer.

[28] (Cuno 1580 ca.) https://daten.digitale-sammlungen.de/~db/0002/bsb00023985/images/

[29] Paolo Giovio writes about a gift by the Emperor Maximilian to the Emperor Suleiman I. Kepler and Giovio refer to different emperors: Maximilian II was the successor of Ferdinand I who died in 1564, while Suleiman reigned from 1520 to 1566, so both Emperors can be the donors of the gift. (Giovio 1552)

[30] Letter to Mästlin, June 1598, line 178-180 (Caspar, Johannes Kepler Gesammelte Werke 1945) p.223

After this overview, Kepler writes about his desire to build a machine that could illustrate the motion of the planets and their proportions, but not with pointers on clock dials, rather with the planets positioned in reference to the zodiac signs, as Archimedes did [32]. In this representation one could put a ruler from the earth to a planet and to the zodiac circle to measure the angle with the accuracy of one degree. Among the list of requirements of this opus, Kepler puts the possibility to move the machine so that one can observe its evolution during a year, with the planets sliding past each other, allowing to compare their motion. The machine should move in the same way as we observe the planets in the sky, either following the natural time (with a clock), or move quickly even for a hundred years and more, to make evident the precession of the equinox. Kepler is aware that this opus will be very expensive, considering also that the construction is expected to work up to a thousand years! And he has just started to build the silver spheres. In the list of requirements, he does not neglect to implement, beside the average motion, also the anomalies of Venus and Mercury and their motion in latitude. He understands how difficult is to reproduce the correct proportions, and he rightly recalls that it is better to limit oneself to imitate the nature, «*natura imitari quantum sufficit*»[33]. The opus will include, anyway, the meridian and the horizon. Kepler's aim is to avoid ornaments in excess, trying to mimic the nature. Concluding, he considers to move each planet with only two wheels. The daily clock will be constructed by a mechanical master. All the wheels will be assembled under a supporting foot.

*VIRTUAL RECONSTRUCTION: A HYPOTHESIS*

*The first concept: a Cup with Platonic Solids*

Kepler's initial concept was to build a silver cup to illustrate the concept of the platonic solids, as in *Mysterium Cosmographicum*, that he calls *Kredenzbecher*, a table decorative cup to contain liquors, beer and wine. In the drawing attached to letter 28 (1596) to the Duke, Kepler identifies the different liquors for each sphere: Sun *"Aquavite"*, Mercury *"Brantwein [distilled wine or brandy]"*, Venus *"Meth" [?]*, Mars *"Vermouth"*, Jupiter *"White wine"*, Saturn *"Other wine or beer"*. It is questionable if the final object would be including the platonic solids or only the planetary spheres; it was conceived to be used as a fashionable and artistic creation for the Duke. I have created a three-dimensional simulation of a possible setting of the spheres and the solids, similar to the *Mysterium Comsmographicum* illustration, to determine the size of the spheres for the planetarium.

I summarize in the table 2 the edge size of the solids and the radius of the inscribed and circumscribed spheres, normalized to the innermost octahedron unitary edge.

| Polyhedral | Edge | Radius | | Ratio |
| --- | --- | --- | --- | --- |
| | | inscribed | circumscribed | I / c |
| Octahedron | 1,00 | 0,41 | 0,71 | 0,57735 |
| Icosahedron | 1,99 | 1,50 | 1,89 | 0,79465 |
| Dodecahedron | 1,70 | 1,89 | 2,38 | 0,79465 |
| Tetrahedron | 11,65 | 2,38 | 7,13 | 0,33333 |
| Cube | 14,27 | 7,14 | 12,36 | 0,57735 |

*Tab. 2 – Parameters of the Platonic solids*

Kepler in *Epitome Astronomiae Copernicanae* [34] publishes a table with the ratio of inscribed and circumscribed radii normalized to 100.000:

---

[31] (Kepler, Epitome Astronomiae Copernicanae 1618) p. 26. It is worth to note that most authors considered by Kepler are not mechanics or horologist: Ramus was a geographer, Giovio was interested to the exploration of newly discovered lands, Cythreus was a theologian. Only Cuno was a mathematician and astronomer, and Turriano (Jannello Torriani) was and engineer and constructor of automata. It is surprising that Kepler did not consider Jost Bürgi, who from around 1580 to 1592 already constructed planetary globes that were considered awesome works of art.
[32] Letter to Mästlin, June 1598, line 195-270 (Caspar, Johannes Kepler Gesammelte Werke 1945) p.223-225
[33] Ibidem p. 225
[34] Ibidem, p. 273 (p. 468 of the manuscript)

|   |   |   |
|---|---|---|
| In Cubo | 57735 | Potestate tertia pars radij circumscripti. |
| Tetraedro | 33333 | Pars tertia radij circumscripti. |
| Dodecaedro | 79465 | Pars ineffabilis, inter duas tertias et tres quintas potentiae radij circumscripti, ablatâ scil. potentiâ |
| Icosaedro | 79465 | Apotomes ab vndecim quindecimis potentiae radij. |
| Octaedro | 57735 | Potestate tertia pars radij circumscripti. |

*Quae sunt istae proportiones orbium in singulis figuris?*
*Semidiameter circumscripti sit 100000. erit inscripti proportio ista.*

In figure 4 the three-dimensional reconstruction of the layout of the spheres and platonic solids.

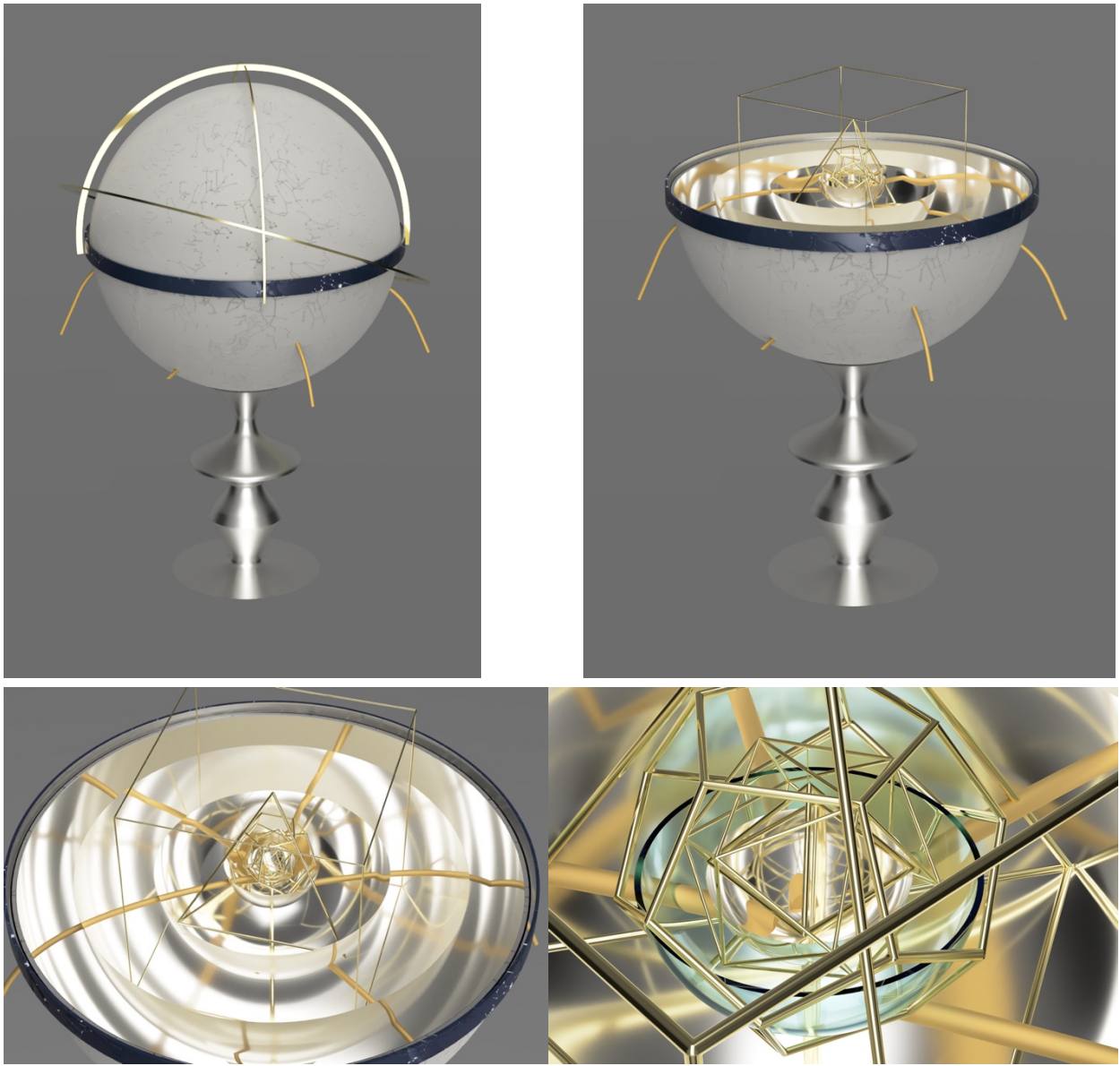

*4 - Views of the reconstruction, with the tubes to spill the liquors. Details of the Platonic Solids.*

This object was never built, as the outer hemisphere would have had to be too large for the inner one to be of a reasonable size.

### *The second concept: a Planetarium*

In the letter 42 dated May 28$^{th}$ 1598, to the Duke Friederich I von Württemberg (1557–1608), Kepler writes that he will construct a *globus* with a planetarium. It is possible and likely that the Duke had in mind a sky globe similar to those already constructed by Eberhard Baldewein (1525-1593), Gerhard Emmoser (1556-1584) or Jost Bürgi (1552-1631), but none of these authors were cited in Kepler's overview. These machines were designed and constructed to show the overall motion of the sky, to identify the position of the stars, to show the motion either of the Sun, the Moon or both. Kepler, in turn, had a clear idea to construct something to show the motion of all planets, and not just the motion of the sky, of the Sun or the Moon. About the he writes in the project description: «*Nothing to say about the moon, since its motion is uniform*» and he did not describe the implementation of Moon's motion: there are no gears for the motion of the Moon or rotation of the Sun. We can assume that one turn of the 60 teeth wheels of the Earth should be equal to one day so that the annual motion is directly controlled by the Earth wheels.

Kepler's design was different from the celestial globes constructed at the time, it is rather a planetarium, where the planets would appear as *standing out*, rotating at their proper distance inside their orbs. While the general concept is clear, its detailed description is disorganized, the sketches are not clean technical drawings; when the description becomes muddled the author writes: «*non possum pingere, tu intellige*», «*I can't draw, you must understand*».

On the other hand, the idea of concentric tubes is very clear: the sketch 5 gives an immediate understanding of the general setup. This sketch shows the concentric tubes and their connection to the driving wheels and to the curved arms that support the planets. The inscription «*linea cubi*» confirms that he used his platonic solids theory to correctly define the planetary distances and sphere measurements.

The planet (in the sketch it is Saturn) moves between two surfaces. In chap. XIV of *Mysterium Cosmographicum* Kepler writes: «*Igitur ut ad principale propositum veniamus: notum est, vias planetarum esse eccentricas et proinde recepta physicis sententia, quod obtineant orbes tantam crassitiem quanta ad demonstrandas motuum varietate requiritur*»[35]. In other words: «once accepted that the path of planets (*vias planetarum*) is eccentric it is also accepted the opinion of physicists that the thickness (*crassitiem*) of the spheres must be sufficient to contain variation of the planet's motion [as required by the equant and deferent]». With this observation Kepler modifies the idea of his predecessors: physicists used to think that the space between the orbs was not empty, Kepler considers each orbs as two hulls within which the planets move. The space between the various orbs is empty and its extension is determined by the eccentricity, while the size of the external hull is determined by the platonic solid.

The exterior of the whole globe is made of silver with the engraved stars. The zodiac is supported by 4 pillars with the four Zodiac symbols ♈♋♏♒. The orientation of the globus is not described but we can assume that the motion axes are vertical and oriented towards the celestial pole, therefore the zodiac circle will be horizontal and the ecliptic circle will be inclined 23.5°. The globus can also have armillary circles, and at least the zodiac ring with engraved degree sign, possibly with subdivision to measure prime and seconds. The mechanism is based on a series of gears whose number of teeth was calculated by Kepler, who writes to have many numbers and he chosen those that are the best for his purpose. We do not have any information about the mathematical method used to calculate these numbers. We know that the first to use the method of continued fractions to approximate a rational number has been Christian Huygens, one century later. In his letter Kepler calculates also the accuracy of his approximation in terms of degrees after 10.000 years; in Table 3, I give the speed ratio computed from the teeth numbers.

| Planet | Wheels | | Ratio |
|---|---|---|---|
| | driving | driven | |
| Mercury | 191 | 46 | 87,97 |
| Venus | 395 | 243 | 224,70 |
| Earth | 60 | 60 | 1 |
| Mars | 42 | 79 | 687,02 |
| Jupiter | 29 | 344 | 4332,62 |
| Saturn | 11 | 324 | 10758,27 |

*Tab. 3 – Wheels*

First of all, Kepler writes: «… *the wheels A will be fixed to their axis and the wheels E will reverse rotate free [on their tubes] (in this way wheels in E have reverse rotation with respect to wheels A)* ». This means that there is a single axis on which all the driving wheels (A) are fixed, while the driven wheels (E) are mounted on a series of concentric tubes. From a construction viewpoint a decision is necessary: weather the wheels are assembled on two parallel direction or along an orthogonal direction. Sketch 2 and Kepler's description suggest first an orthogonal assembly: in line 414 p. 172 he writes « *NB the wheels in A are not vertical, on the contrary they are aligned horizontally and perpendicular to CE* ».

---
[35] (Kepler, Mysterium cosmographicum 1596) p. 47

In Figure 5 we see how orthogonal or parallel wheels can be matched: on the left the horizontal wheel and the vertical one ("crown wheel") match at a distance that depends on the diameter of the pitch circle (Figure 6) of the first wheel and the outer diameter of the second. The configuration to the right implies that the two wheels match when pitch circles are tangent. In any case every gear couple must have the same pitch. To determine the size of the wheels we must recall the fundamentals of horological gear theory: the primitive (or pitch) circles of two wheels must be tangent for a smooth rotation. The pitch circle and the teeth number are related by the *module,* i.e., the ratio pitch_diameter/teeth_number. The pitch diameters will vary in a wide range given the very different teeth numbers; therefore, a constant module for all the wheels cannot be adopted. On the contrary, the modules must be chosen to get a constant distance between the centers of each couple of wheels. A possible size and pitch of the gears assembled on parallel axes are summarized in the table 4. Assuming the units are millimeters, we have a smallest wheel diameter 11mm and largest wheel 309 mm.

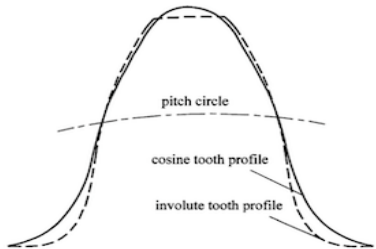

*Figure 6 - The pitch circle*

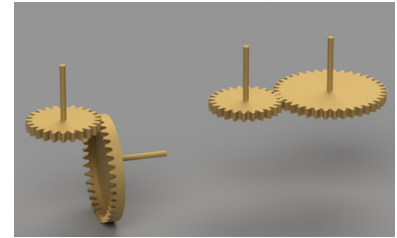

*Figure 5 - Orthogonal and parallel gearing*

| **Wheel couples** | **Module** | **Primitive diameter** | | **Center DIstance** |
|---|---|---|---|---|
| | | **Driving** | **Driven** | |
| **11 - 324** | 1,00 | 11,00 | 324,00 | 167,50 |
| **29 – 344** | 0,90 | 26,10 | 309,60 | 167,85 |
| **42 - 79** | 2,76 | 116,00 | 218,20 | 167,10 |
| **60 -60** | 2,79 | 167,40 | 167,40 | 167,40 |
| **395 - 243** | 0,53 | 127,58 | 207,38 | 167,48 |
| **191 - 46** | 1,41 | 269,31 | 64,86 | 167,09 |

*Tab. 4 - Wheel size and module*

An orthogonal assembly, following Sketch 1 and 2, a solution could be as in Figure 7 (left). This solution is very difficult to construct for the irregular distribution of the weight and the size of the whole machine. A feasible assembly would be to organize gears along two parallel axes, even if this is not what Kepler first wrote. Later, while describing the ground motion, we understand that he adopts the parallel axis assembly.

To cut the teeth is a technical problem: a wheel with 191 teeth driving 46 teeth requires a relatively strong torque; moreover, to divide a circle into 395 parts it is a difficult task and in general large teeth number are difficult to cut. Moreover the couple 192-46 need a very high module number: 12,2. Kepler is not fully aware of the difficulties of this construction; he only notes the difficulties related to gears with small radius. Anyway, Schickard and Kretzmeyer raised the issue with Mëstlin and the Duke.

*The spheres*

The discussion on the weight of the spheres reveals the adherence to the homocentric sphere theory, as already said, but it is unclear how the spheres should be positioned.

Kepler writes that « *The load is on E as much as you want, nevertheless it weights but separately, as we will see. And if the weight is mutually on each other, nevertheless they will be light once removed the bodies and the orbes.*».

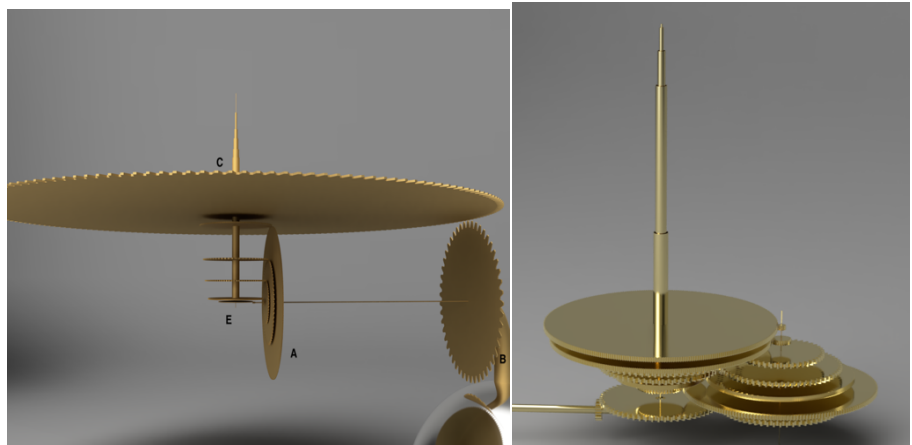

*Figure 7 - Orthogonal assembly of the gears. Letters identify positions in Sketch 1 and 2. Left: orthogonal setting. Right: parallel setting*

Further « *I have already written that the opus C, suspended in its center to the external part of the fixed stars globe, with, try to understand, balanced supports, maybe difficult to build. C is driven by E, that rotates (reversed) around B, preserving the fixed parts. Similarly, BE in A is bend so that while AE and EB match, A makes B rotate concentrically and E eccentrically I said that the hole for B shall not be in the zodiac pole, but at a distance large as the size of AE. This distance determines the size of the work, how to build the wheels as small as is possible.* » Later he writes that the *orbes* (opus C) are solids and divided in half, not pierced. We can conclude that Kepler speaks of hemi spheres, that allow to observe the interior and the motion, therefore the upper hemi sphere could be transparent or removed. Moreover, the whole opus rotates around B, not around the zodiac pole.

A question arises: how the planets are positioned and how do they move with respect to the spheres? First of all, Kepler writes: « *Part C is driven so that the earth remains at the center of the orbis of fixed stars*». This means that his idea was not to show the motion of the planets around the Sun, rather to show how the motion of the planets around the Sun is seen from the Earth. He continues: «*It is evident that if the motion in C follows a path HIK, it is necessary an opposite motion KIH that keeps it in the center [Sketch 4]. Because if the earth moves correctly also all other planets will move accordingly in the sky, and together with the earth*». This point is further analyzed in the last part about the *primum mobile*. Kepler writes: « *Now let's come to C. It will be made with orbibus divided in half and without holes. They will be fixed to the bodies so that no orbis will be movable* », and from this we could conclude that the spheres will stay fixed and the planets will move inside. But fixed to what bodies, «*corpora*»? Kepler draws the sketch 6 and writes: « *I presume that it will not be too difficult if the machine is extended beyond the half globe and the planets will pass, like under a yoke, under three or 4 small rods.* » In my interpretation this will be solved with the bars that sustain the spheres (figure 8) that are modified so that the planets can pass under them.

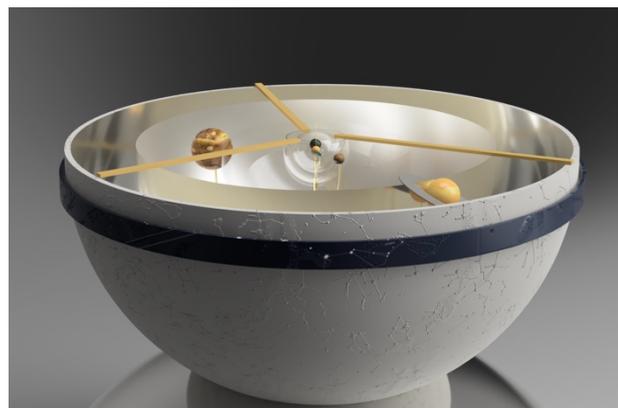

*Figure 8 - Bars to sustain the spheres*

For the variation from aphelion to perihelion Kepler has a simple and clever solution: the external surface of the orbs is deformed into a convexity and an opposite concavity, while the arm of the planet will be flexible and elastic to slide on the surface and be pushed towards the center at the aphelion and pulled out at the perihelion. The deformation should be set coherently to the direction of the apsidal line. I have modelled this deformation only for the planet Saturn.

The middle hemisphere, in the figure 9, is the circumscribed sphere of the platonic cube, the other two spheres delimit the thickness for the Saturn orbit and are eccentric to the central axis. Kepler describes how he would implement the eccentricity of the orbits and writes an obscure description: *«Ich will mich drüber setzen und raitten (quandoquidem ipsum meum inventum, quod par est, in opere exprimi, calculum anteà mutat) ob es vil austrage, wan ich ubique 1/10 diametrj pro eccentricitate näme »* *« I want to determine and fix the eccentricity, if I can, to 1/10 of the diameter everywhere (since in my own invention, as I have described, the calculus changes what is equivalent).*

To implement the variation in latitude Kepler suggests to put under the hull of Saturn a surface inclined as its orbital plane. I could not figure how to implement this solution keeping the hemispheres of the orbs. The solution adopted in planetary machines constructed from the XVIII century had a tilted orbital surface, over which an axis slides varying the latitude of the planet; the description by Kepler looks similar to this solution.

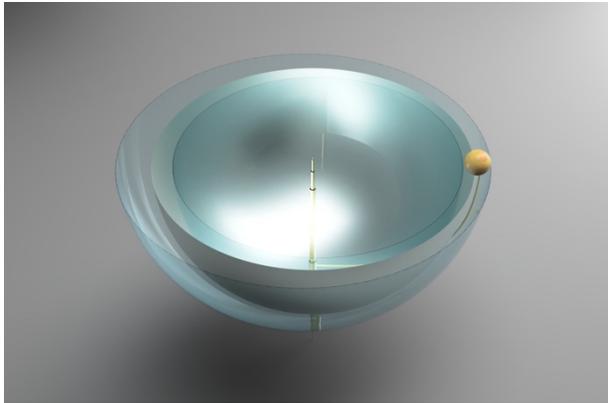

*Figure 9 - Thickness and ovalization of Saturn orbis*

Summarizing: each orbis is composed of two surfaces, they are fixed and oval shaped to implement eccentric motion. They have a hole in the center to allow the passage of the concentric tubes for the motion of the planets. The outermost sphere is the fixed star and it is sustained by the basement of the globe, the other spheres are sustained by three or four bars. To keep the bars sufficiently elevated to allow planets' motion, the spheres are modified to extend over the orbital planes. This interpretation solves the above seen problem: Kepler affirms that the spheres do not move, but if each sphere is supported by the corresponding concentric tube they will be forced to rotate with the planet. On the other hand, if the spheres are suspended with three bars, they will remain fixed and the planets will be allowed to rotate and slide on the sphere surface as required for the first anomaly.

### *Primum motum*

This is the core of Kepler's concept, drawn in sketch 7. Kepler writes a long description for this device that he calls *primum motum* and later *ground motion*. In his view it should reproduce the Copernican system as seen from the earth.

Let's recall what Kepler writes, while observing the scheme in figure 10.

«... *in point B let be an axis BA that rotates around the fixed-point A and keeping CD always parallel to itself. The point B on the bar CD describes a circle G with center A, while others have different centers, e.g.,*

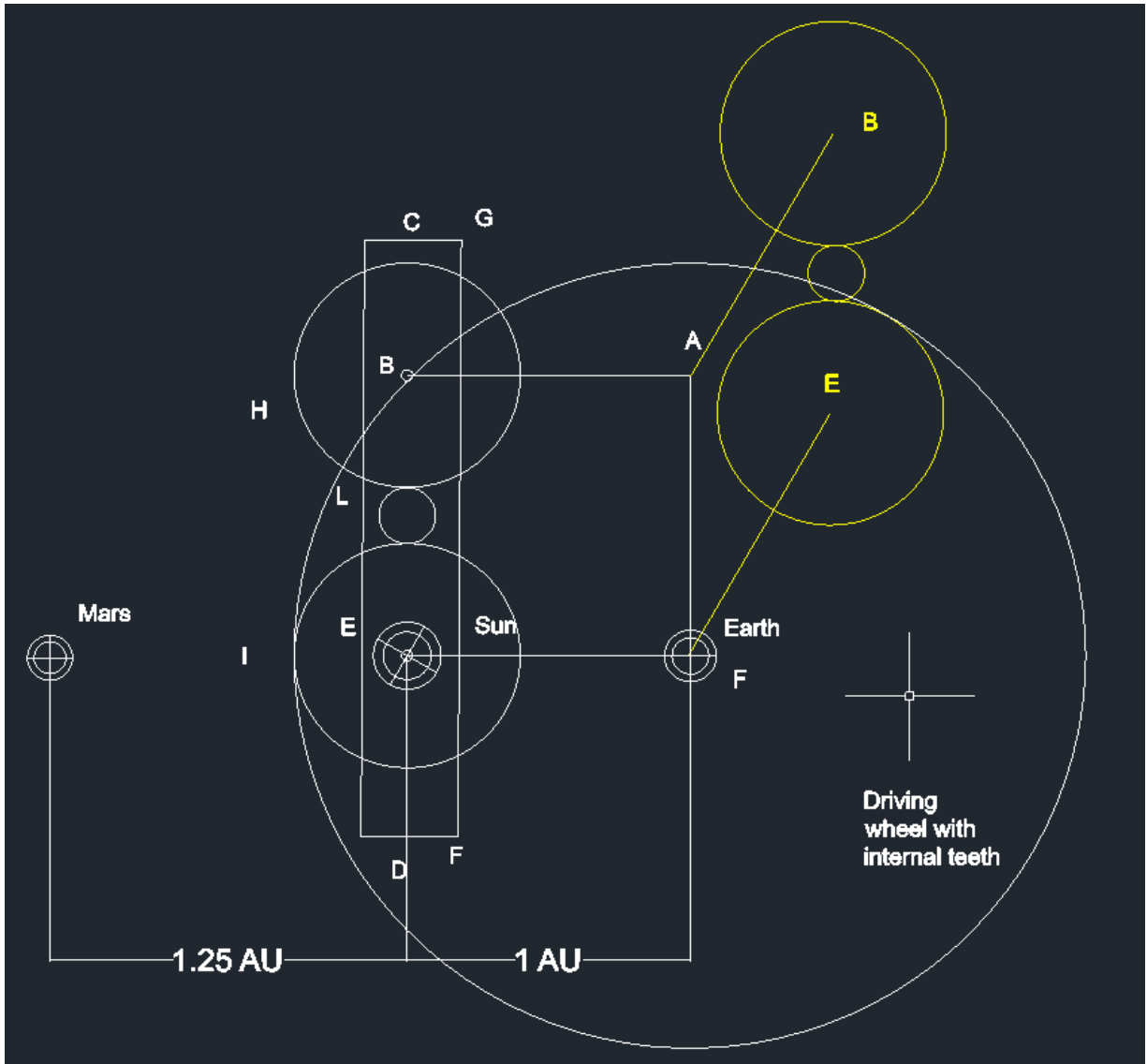

*Figure 10 - Kepler's scheme, with the driver (the ring with internal teeth) the Sun, the Earth and Mars. Mars distance respects Kepler's platonic solid theory. In yellow the Ab, EF wheel H and I after rotation of 120° clockwise*

*E describes a circle with center F».* By this description the bar CD rotates parallel to itself so that points B and E describe circles centered in A and F.

The motion of this mechanism is described by Kepler:

*Thus the circle around F led by E is equal to the orbis of the Earth and it is as if F were led along a circle around E. Now imagine 6 motor circles pivoted in B, and passing through a hole in the rod CD. In E there are 6 free wheels adapted to the above. Now let B rotate towards G, then E will rotate towards A and H will rotate on the axis towards G (NB: it would be better to reverse the whole motion from right to left.) Let there be a circle around E led by the adjacent circle in the opposite direction towards I and D. Whatever the motion of the lower wheel, such will be that of the upper planet which is represented in KF[36], and that is the earth. In this situation F is the Earth, E the position of the Sun. An intermediate wheel L reverses the motion so that the Sun in E moves towards A, H moves towards C, point I towards B and F towards D. The Earth is and remains in F. This is also the case for the other planets. They, because of the different pairs of wheels in B and in E, will move with different velocities. Thus I will move 6 planets, 12 wheels and 6 actors. The length of BA can also be adjusted at the extremes C or D, so that there is enough space for wheels of many teeth.*

---

[36] Letter K does not exist, read as HF

In this description there is a problem: the rotation direction must be anti-clockwise, so that the rotation of I moves correctly the Sun around the Earth. But the presence of the intermediate wheel L forces H to rotate in the same direction, as well as the driving wheels fixed on the B axis, therefore the wheels pivoted on E will drive the planets clockwise.

As a consequence, the wheels I and H must rotate opposite each other, and wheel L must be removed. Moreover, the wheels that move the Earth in the concentric-tubes assembly should also be removed, otherwise the Earth would rotate anti-clockwise around E moving it from the desired position F. The solution I propose is shown in figure 19, right, where the wheel L and the driving and driven wheels of the Earth are removed, so that the Earth is subject only to the rotation around A generated by the ring wheel. To activate the rotation of wheel I, a second identical wheel is fixed to its same axis and below the bar CD; this wheel is moved by another wheel, that could be shaped as a ring with internal teeth.

Now assume that the Earth is positioned along the F axis, assume moreover that the ring wheel is fixed in center F, therefore its anti-clockwise rotation has the effect to move wheel I in the same direction and the bar CD parallel to itself, so that the whole machine, inside the sphere of the fixed star, rotates around the center F, and, as said by Kepler, the Earth remains on F and the Sun rotates anti-clockwise around it. Moreover, wheel I transfers an inverted rotation to the wheel H and to the axis to which all the driving wheels of the planets are fixed, and each of them transfer the motion clockwise to the planets, excluding the Earth. But in this case the Earth is subject to two motions: one around F for the rotation of the ring wheel and one around its own axis for the rotation of its driven wheel. Therefore, the Earth will not keep the central position on F. To avoid this, Kepler considers the effect of the H, I and L wheel chain the rotate clockwise the Earth, keeping it on center F.

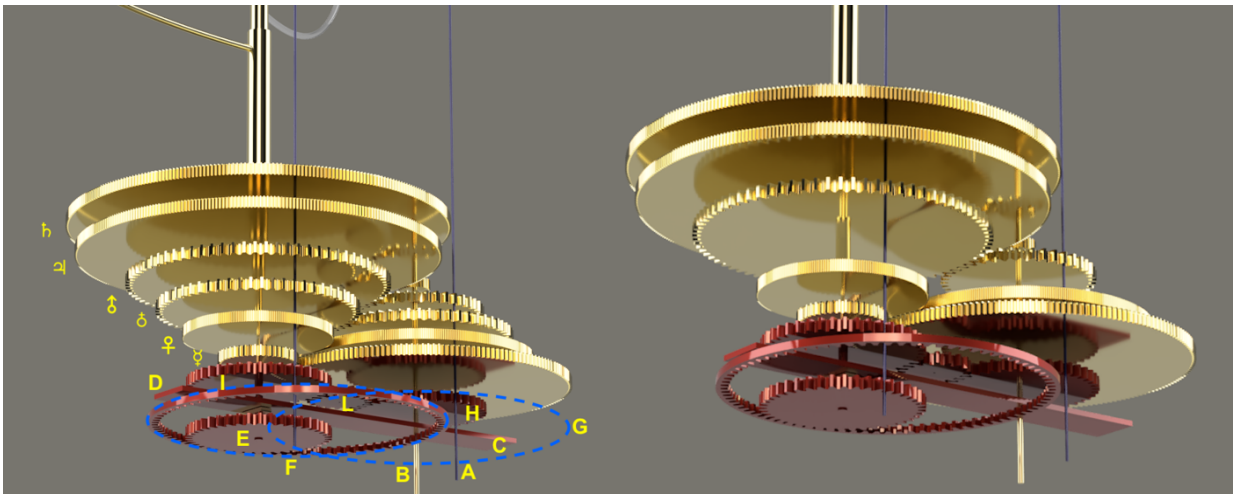

*Figure 11 - Left: Kepler's description. Right: feasible implementation. The light blue mark centers A and F of Kepler's sketch*

This motion produced by the rotation of the bar CD parallel to itself, makes the whole solar system to rotate around the center F, where Kepler keeps the Earth fixed. We get finally a configuration of the motion that emulates a Tychonic model, allowing to show the motion of the mobile stars as seen from the Earth.

In the figure 12 one can see a graphical simulation of the retrograde motion of the planet Mars.

The loop of the retrograde motion is quite different from the real one, this because of the size of Mars orbit radius: by Kepler it is 1.259 times that of the Earth, based on the theory of platonic solids, while the real radius is 1.52.

The fixed star sphere will be much larger to host this complex mechanism, and will by fixed and centered on the Earth, while the orbs rotate with the ring wheel and bar CD, and the planets' motion will be the combination of this motion and their own rotation around their common axes centered on the Sun.

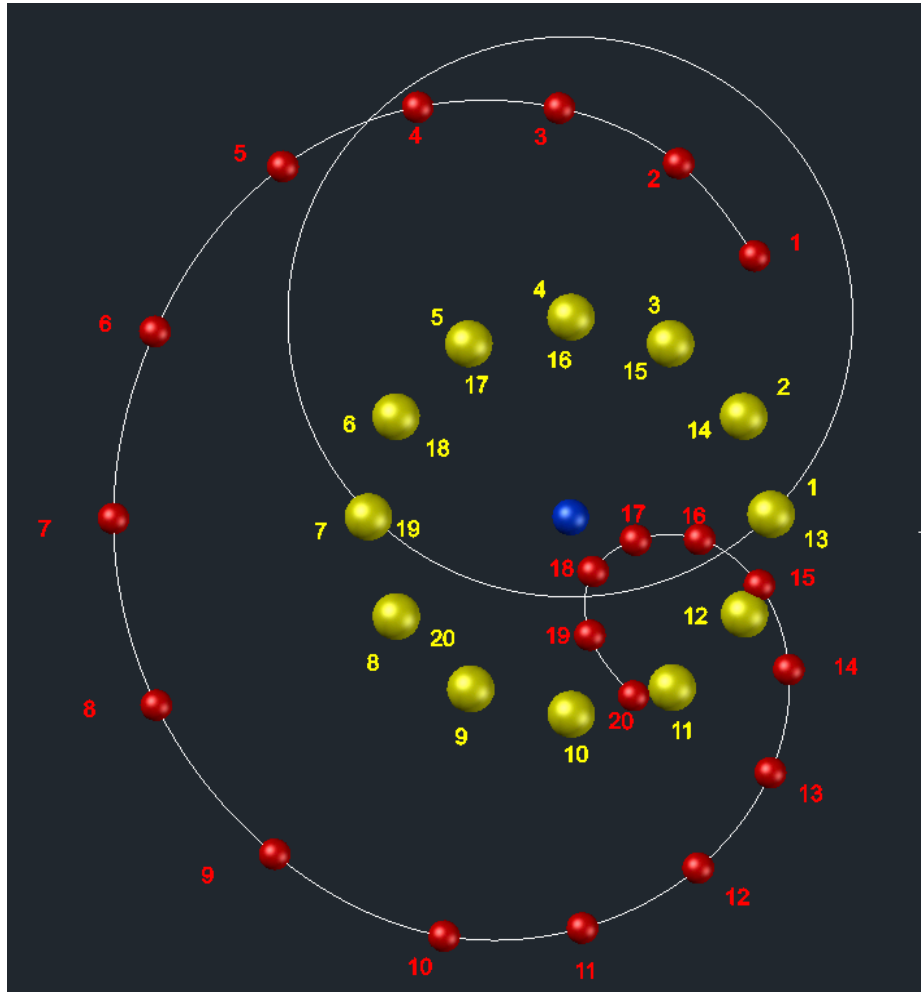

*Figure 12 - Retrograde motion of Mars (red). The numbers denote the correspondence between the position of the Sun (yellow) and of Mars.*

The configuration of Kepler's planetarium is very original: it allows an observer to see the solar system from above, like flying in the space. While he was computing Mars retrogradation motion and drawing what he called *panis quadregesimalis* in his work *Astronomia Nova,* it is likely that he was considering this observation point for the transformation of the observed coordinates of the planets collected by Tycho and his pupils. The interpretation by Prager (two models into a single machine) is correctly implemented in my reconstruction as the motion of the planets is considered as observed from the Earth, in the center of the machine, while all moving bodies rotate around the Sun.

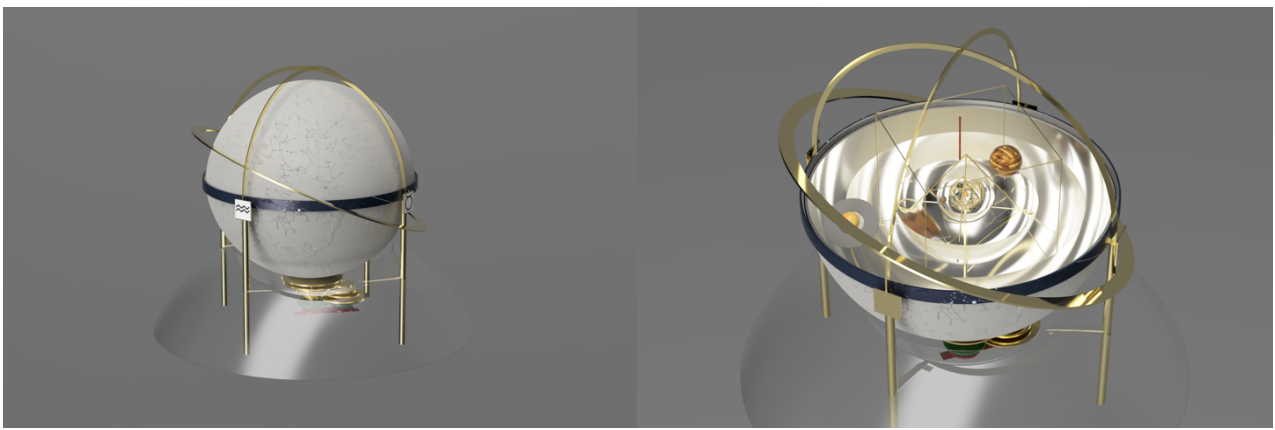

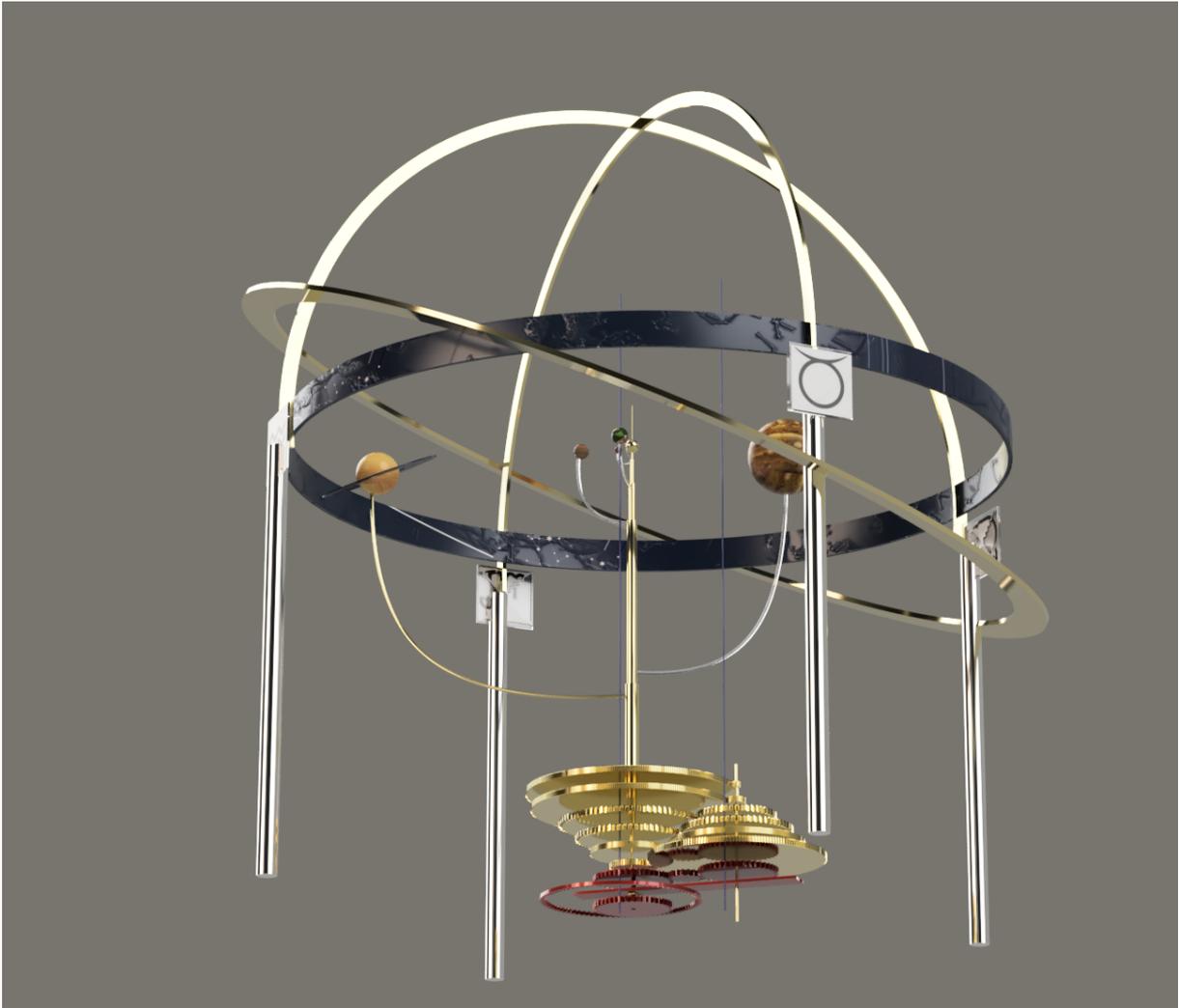
*Figure 13 - Final rendering of the hypothetical reconstruction with and without the platonic solids and Spheres*

In figures the final rendering of the reconstruction. I have also inserted the colures, the zodiac (in blue) and the equator to clarifies Earth axis orientation. The globe is divided in two parts, the top one can be removed to observe the interior of the planetarium.

## *CHRONOLOGY OF THE WORK*

In a series of letter, beginning in February 1596, when it was first conceived, until April 1599 when the work was abandoned, we can read the difficulties of this work and the evolution of Kepler's concept, from a cup to expose the theory of platonic solids to a planetarium.

### 1596
- *February 17th letter 28.* Kepler writes to the Duke Friederich I von Württemberg. Kepler, having left Graz to his Swabian homeland, proposes to the Duke to allow him to build a model shaped as a table cup, a *Kredenzbecher*, to expose the structure of cosmos as described and drafted in the book *Mysterium Cosmographicum*. Kepler sends to the Duke a paper model (a drawing) of the table cup. He will request an opinion to Mästlin before proceeding. (see Figure 5 )
- *March 12th letter 31.* Mästlin writes to the Duke defending Kepler's theory about the distances among the planets, that are also coherent to astronomical observation.
- *April 13th letter 38.* Kepler writes to Mästlin that the construction, that he now calls *opus argenteum,* is proceeding slowly: «*Opu argenteum ita tarde procedit …*».
- *May 28th letter 42.* Kepler writes to the Duke Friederich, that he dismisses the work on the table cup, to work on an «*Astronomicum Opus*».

- *June 27th letter 49.* Kepler writes to the Duke that he instructed a goldsmith to work on the new project, that he calls now «*Astronomischen Werk*».

**1597**

- *April 9th letter 64.* Kepler to Mästlin. After a discussion about the calendar reform by Pope Gregorius, clarifies the conception of platonic solids commenting tab. 76 of his manuscript and adding this table:

| Sicut se habet circum-scriptus | Cubo / Tetr. / Dod. / Icos. / Oct. | ad inscrip-tum: ita se habet | Media motoria Saturnj ad mediam motoriam Jovis. / Ima ♃ Copernicana ad sum-mam ♂ Copernicanam. / Ima ♂ Copernicana ad altis-simam terrae cum Luna Cop. / Ima ⊕ simplicis Copernicana ad mediam motoriam ♀. / Summa ♀ Copernicana ad mediam ☿ motoriam. | Quàm pro-ximè |

- *April 27th letter 67.* Mästlin writes to Kepler that he will intervene, since the goldsmith has done an ugly work «*Stellae non eleganter exsculptae sunt*» "The stars are not elegantly engraved" and what has been already done is not enough.
- *July 11th letter 71.* Mästlin writes to Kepler that the execution of machine is not clear. The internal mechanism raises doubts, unsuitable for a high-quality work.
- *Beginning of October, letter 75.* Kepler writes to Mästlin about the "unhappy silver work", suggesting to give the work to a clockmaker in Graz.
- *October 30th letter 80.* Mästlin writes to Kepler that the difficulties of the construction of the machine become more and more evident. He writes that the many doubts about the work are now evident, and the objective is now to avoid to upset the Duke.
- *December 24th letter 83.* Kepler writes to Herwart von Hohenburg in München. Kepler alludes to the Planetarium and the solution he found using few wheels for the motion of planets with an accuracy up to 6.000 years.

**1598**

- *January 6th letter 85.* This is the letter containing the detailed description of the project.
- *March 15th letter 89.* Kepler writes to Mästlin that he will let the clockmaker to continue the construction when the needed silver will be available. «*Transeo ad Penelopes telam. Si mihi illa massa argentj daretur unde automatarijs suppeditare possum, instruirem horologium caeleste, postea tantum filium ad repraesentanda corpora circumducerem. Esset autem illud tale, ut si ex polo eclipticae quis inspiceret, vederet planetas in suis signis*». "I come to Penelope loom. And I remain in my first opinion. If I receive an amount of silver to accelerate the work of the constructor, I could set up the celestial clock, so that after so much work we would arrive to represent the bodies. The work would be such that if somebody looks from the ecliptic pole, he could see the planets in their zodiac signs".
- *March 11th letter 90.* The duke Friederich writes to Kepler that what has been made until now is not enough. The Duke wants that the work be done again, and demands Kepler to prepare an entirely new plan. The critics are about the quality of the engraving of the stars on the globus: «*Die Stern daruf recht gestochen seien*», "The stars must be right engraved".
- *May 2nd letter 97.* Mästlin (whose child Augustus died the February 10th for epilepsy) writes to Kepler to make new proposals for the machine: «*... you shall declare and describe what you have in mind about the orbits disposition, about the motion, and first of all how to build...*»
- *June 1st / 11th letter 99.* Kepler answers to letter 97 clearing the requirements and reviewing known astronomical clocks and machines (see above).
- *July 4th letter 101.* Mästlin writes to Kepler to report on the work and that he has some concerns about the technical details.

**1599**

- *January 11th/12th letter 110.* Mästlin writes to Kepler that there are other problems: the goldsmith has no more interest.
- *February 16th/26th letter 113.* Kepler writes to Mästlin that he hopes to abandon the work to protect his reputation to the Duke.
- *April 12th letter 119.* Mästlin writes to Kepler that the work is abandoned.

It is worth noting that the major critics and concerns by Mästlin and the duke Friederich were not on the mechanism and its construction problems, but on the quality of the work of the goldsmith.

## CONCLUSIONS

When Kepler proposed the first concept, he did not have the ample vision that he gradually developed. The difficulties of the construction were related to practical problems with goldsmiths or mechanics experts rather than in the astronomical concept. On the other hand, since the *Kredenzbecher* had the purpose to contain in the spheres wine or liquor, their size was not adequate: the interior orbs were too small to contain sufficient liquid. So, a failure of the project would have undermined Kepler's invention. With the second concept Kepler affords a problem that is astronomical in itself and includes not only the invention of the platonic solids to define planets' distances, more important it includes the idea to display with a machine, for the first time, the motion of the planets as caused by a single motor, while focusing on the apparent motion. This project was also a failure, but Kepler traced a pathway for future mechanical inventions for the construction of planetary machines.

The project of Kepler is an anticipation of his major innovative approach to the study of the sky: to move outside of the Earth, flying with his imagination high in the space to observe the motion of the planets and to describe their paths, what he did while computing Mars retrograde motion and plotting the drawing of the *panis quadragesimalis*[37].

An effective way to explain this concept can be done today with computational methods. An example is the simulation of Kepler's planetary configuration implemented by Paolo Maraner[38] using Mathematica ™.

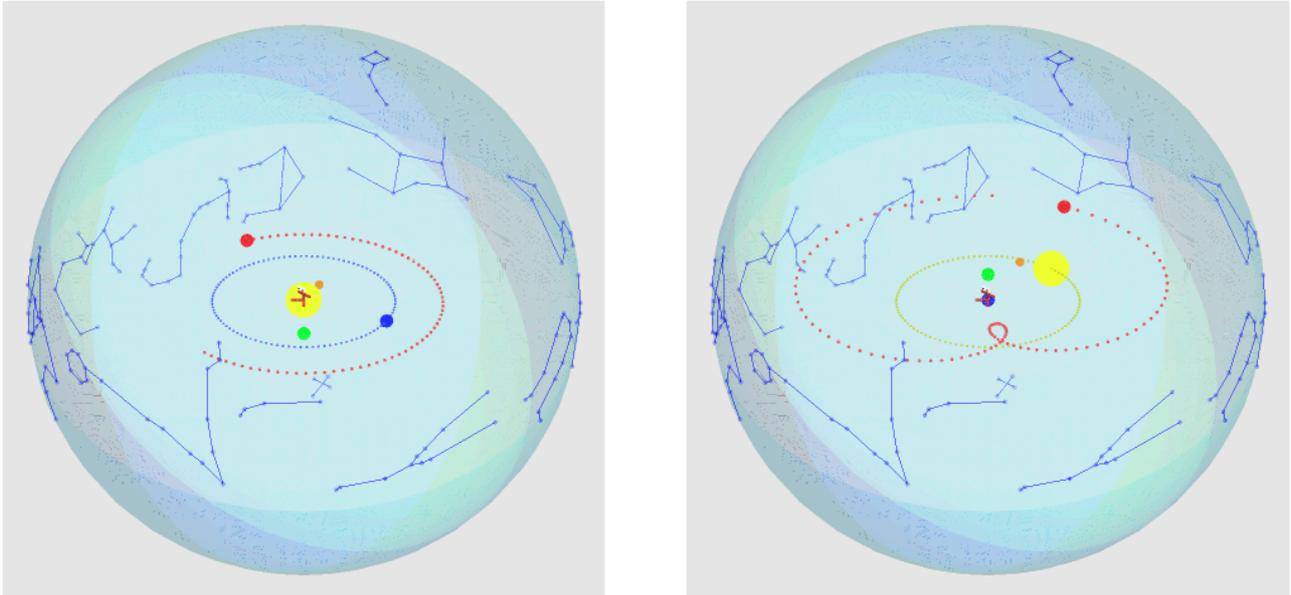

*Figure 14 – Computational simulation. Viewpoint: left heliocentric; right geocentric. In both cases the motion is heliocentric*

Kepler's theory of platonic solids to determine planetary distances gives an estimate of 1.259 astronomical units (AU) of Mars - Sun distance, while the correct value should be 1.52 AU, with Earth-Sun distance equal to 1 AU. To evaluate the effect on retrograde motion as the planet's distance from the Sun varies, the result of a 15-year simulation can be seen in the figure, computed with Matlab™. Diolatzis et al. [39] simulated the geocentric motion of Mars, showing the correspondence between the computed path with the Kepler diagram; they adopted the actual distance of Mars from the Sun of 1.52 AU, instead of that derived from the theory of Platonic solids.

---

[37] (Kepler, Astronomia nova seu Physica coelestis: tradita commentariis de motibus stellae Martis, ex observationibus G.V. Tychonis Brahe 1609)
[38] Ptolemaic to Copernican World System Continuum http://demonstrations.wolfram.com/PtolemaicToCopernicanWorldSystemContinuum/
 Wolfram Demonstrations Project, Published: July 13, 2017
[39] (Diolatzis e Pavlogeorgatos 2019)

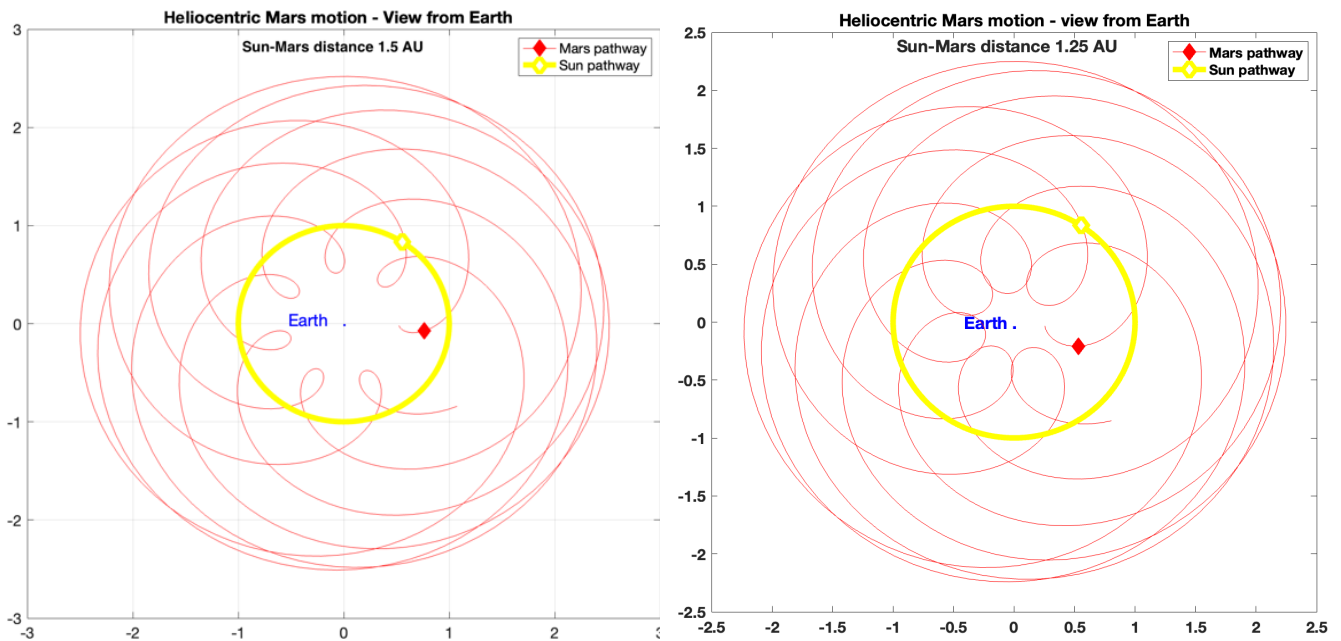

*Figure 15 - Difference of retrograde motion as Mars distance varies from its actual value and the value computed from Platonic Solid theory*

For the first time the solar system is conceived as planets moving independently around the Sun, under the effect of a unique force, the primum mobile, that later Kepler identifies in a kind of magnetic force[40] acting from the Sun to all the planets.

His model does not contain the motion of the Moon and neither the annual motion, therefore it is not an analog astronomical computer like (hypothetically) the Antikythera machine. It is an educational instrument for the lay people and for the scholar to better understand the Copernican theory and the cinematics of planet apparent motion.

Kepler's theory of primum mobile, that he properly calls *ground motion*, is clearly represented in the virtual reconstruction: the planets are moved by a gear system, possibly with a single clock as motor. There is no need of complex artifacts as required for the nested homocentric spheres.

Kepler wants to implement the eccentricity of the motion of the planets by using epicycles. Apart from this sentence at the beginning of the project description, no other details are provided about epicycle implementation. Indeed, the eccentricity of planet motion will be solved by a trick: the deformation of the sphere (orbs) that contains the planet orbit into an oval, that will force the flexible support of the planet's body to bend and move closer or farther to the Sun.

In the virtual reconstruction presented in this paper, I adopted a parallel setup for the series of driving and driven wheels. I left one open problem that has been considered by Kepler without providing a solution: the change of the latitude of planet motion, i.e., their orbital plane orientation. The planetary machines of the XVIII century could solve this problem by allowing the bodies, in particular the Moon, to slide on an inclined surface, modifying in this way the latitude. While suggesting this same solution, I think it would require major changes, since the planets are rotating inside hemi spheres that should be cut into spherical ring to allow the passage of a vertical rod sliding on an inclined plane.

REFERENCES

Andrews, N. «Gilding Kepler's cosmology.» *Journal for the History of Astronomy*, 2021: 3-32.
Caspar, M., a cura di. *Johannes Kepler Gesammelte Werke.* Vol. XIII. München: Beckische Verlagbuchshandlung, 1945.

---
[40] (Krafft 2005)

—. *Kepler.* London: Dover. Digital Edition, 1993.
Cuno, J. *Brevis Descriptio artifciosi, novi & Astronomici automati horologici, cuius simile antehac non exstitit, Inventi Primum Srudio & industria, Magistri Iacobi Cunonis Francofordie ad Oderam.* Frankfurt am Oder, 1580 ca.
Diolatzis, I.S., e G. Pavlogeorgatos. «Simulating Kepler's Geocentric Mars Orbit.» *New Astronomy* 71 (2019): 39-51.
Giovio, P. *Pauli Iovii Nouocomensis episcopi Nucerini Historiarum sui temporis Tomus Secundus.* Florentiae: Laurentii Torrentini Ducalis Typographi, 1552.
Goldstein, B. R., e G. Hon. «Kepler's move from Orbs to Orbits: documenting a revolutionary scientific concept.» *Perspective of Science* 13, n. 1 (2005): 74-111.
Kepler, J. *Astronomia nova seu Physica coelestis: tradita commentariis de motibus stellae Martis, ex observationibus G.V. Tychonis Brahe.* Praga, 1609.
—. *Epitome Astronomiae Copernicanae.* A cura di M. Caspar. Vol. 7. Lenti: Kepler Gesammelte Werke, 1618.
—. *Mysterium cosmographicum.* Tubingae: Georgious Gruppenbachius, 1596.
King, H. C., e J.R Millburn. *Geared to the Stars, the Evolution of Planetariums, Orreries, and Astronomical Clocks.* Toronto, 1978.
Krafft, F. «Einleitung - Neue ursächlich begründete Astronomie.» In *Neue ursächlich begründete Astronomie*, di J. Kepler, 1-40. Marixverlag, 2005.
Martens, R. «Kepler: Models and Representations.» In *Johannes Kepler. From Tübingen to Żagań*, di Kremer R.L., & J. Włodarczyk, 239-251. Warszawie: Instytut Historii Nauki PN, 2009.
Mosley, A. «Objects of Knowledge: Mathematics and Models in Sixteenth-Century Cosmology and Astronomy.» In *Transmitting Knowledge: Words, Images and Instruments in Early Modern Europe*, di S. Kusukawa, & I. Maclean, 193-216. Oxford: Oxford UP, 2006.
Prager, F. «Kepler als Erfinder.» A cura di F. Krafft, K. Meyer, & B. Sticker. *Internationales Kepler-Symposium - Weil der Stadt.* Hildesheim: Verlag Dr. H.A. Gerstenberg, 1971. 385-407.
Schickard, W. «Wilhelmi Schickardi autographa - Cod.hist.qt.203.» *Digitale Sammlungen der Württembergischen Landesbibliothek.* 1610. http://digital.wlb-stuttgart.de/purl/bsz411134515 (consultato il giorno April 2022).
Vitali, H. *Lexicon Mathematicum.* Roma, 1559.
von Dyck, W. «Die Keplerbrief auf der Braunschweigischen Landesbibliothek in Wolfenbüttel. II. Teil.» *Abhandlungen der Beyerischen Akademie der Wissenschaften, Mathematisch-naturwissenschaftliche Abteilung*, 1934: 1-88.